\documentclass[aps,pra,amsmath,amssymb,twocolumn,superscriptaddress,notitlepage,reprint]{revtex4-1}
\usepackage[utf8]{inputenc}
\usepackage[T1]{fontenc}
\usepackage{graphicx}
\usepackage{dsfont}
\usepackage{color}
\usepackage{epsfig}
\usepackage{bm}
\usepackage{times}
\usepackage{bbm}
\usepackage{subfigure}
\usepackage[colorlinks,citecolor=red,linkcolor=green,urlcolor=blue]{hyperref}

\newcommand{\m}{\mathrm{m}}
\renewcommand{\c}{\mathrm{c}}
\renewcommand{\a}{\mathrm{a}}
\renewcommand{\in}{\mathrm{in}}
\newcommand{\out}{\mathrm{out}}
\newcommand{\BS}{\mathrm{\scriptscriptstyle{BS}}}
\newcommand{\DC}{\mathrm{\scriptscriptstyle{DC}}}
\newcommand{\SQL}{\mathrm{\scriptscriptstyle{SQL}}}
\newcommand{\CQNC}{\mathrm{\scriptscriptstyle{CQNC}}}
\newcommand{\ZPF}{\mathrm{\scriptscriptstyle{ZPF}}}

\begin{document}

\title{Coherent Cancellation of Backaction Noise in optomechanical Force Measurements}

\author{Maximilian H. Wimmer}
\affiliation{Institute for Gravitational Physics (Albert Einstein Institute), Leibniz University Hannover, Callinstra\ss{}e
  38, 30167 Hannover, Germany}

\author{Daniel Steinmeyer}
\affiliation{Institute for Gravitational Physics (Albert Einstein Institute), Leibniz University Hannover, Callinstra\ss{}e
  38, 30167 Hannover, Germany}

\author{Klemens Hammerer}
\affiliation{Institute for Theoretical Physics, Leibniz University Hannover, Callinstra\ss{}e
  38, 30167 Hannover, Germany}
\affiliation{Institute for Gravitational Physics (Albert Einstein Institute), Leibniz University Hannover, Callinstra\ss{}e
  38, 30167 Hannover, Germany}

\author{Mich\`{e}le Heurs}
\affiliation{Institute for Gravitational Physics (Albert Einstein Institute), Leibniz University Hannover, Callinstra\ss{}e
  38, 30167 Hannover, Germany}

\date{\today}

\begin{abstract}
Optomechanical detectors have reached the standard quantum limit in position and force sensing where measurement backaction noise starts to be the limiting factor for the sensitivity. A strategy to circumvent measurement backaction, and surpass the standard quantum limit, has been suggested by M. Tsang and C. Caves in Phys. Rev. Lett. \textbf{105} 123601 (2010). We provide a detailed analysis of this method and assess its benefits, requirements, and limitations. We conclude that a proof-of-principle demonstration based on a micro-optomechanical system is demanding, but possible. However, for parameters relevant to gravitational wave detectors the requirements for backaction evasion appear to be prohibitive.
\end{abstract}

\maketitle

\section{Introduction}

The accuracy of any quantum mechanical measurement is limited by the Heisenberg uncertainty principle. For a force measurement based on an optomechanical sensor \cite{Caves1980a,Clerk2010a,Danilishin2012,Chen2013,Aspelmeyer2013}, that is a harmonically bound test mass whose position is probed optically, this principle is present in Poisson statistical amplitude quadrature noise at the detection port (also known as shot noise) and radiation pressure backaction noise in the phase quadrature, introduced by the harmonic oscillator \cite{Caves1981}. Shot noise is a well known effect limiting high-precision interferometric experiments such as LIGO at high frequencies \cite{Abramovici1992}. Radiation pressure noise, which was recently observed for the first time \cite{Purdy2013,Murch2008}, will be limiting in the low frequency regime of next generation gravitational wave detectors \cite{Harry2010}. The point at which both of these noise sources contribute equally to the total noise gives a lower bound for classical detection sensitivity and is called the standard quantum limit of interferometry (SQL) \cite{Caves1980a,Clerk2010a,Danilishin2012}.

To overcome the SQL in force measurements, different approaches have been proposed: Frequency-dependent squeezing \cite{bondurant1984} and variational measurement \cite{Kimble2001}, the use of Kerr media \cite{Bondurant1986}, dual mechanical resonators \cite{Briant2003,Caniard2007}, the optical spring effect \cite{Chen2010}, stroboscopic measurements \cite{Braginsky1995a}, or two-tone measurements \cite{Braginsky1975,Thorne1978,Braginsky1980a,Clerk2008,Suh2013a}. Some of these have been demonstrated experimentally \cite{Chelkowski2005,Mow-Lowry2004,Sheard2004}. A novel back-action evading scheme suggested by Tsang and Caves \cite{Tsang2010} proposes to coherently cancel out the effects of backaction noise; this is dubbed coherent quantum noise cancellation (CQNC). In contrast to other proposed backaction-evading techniques, this scheme destructively interferes the backaction noise with its counterpart, an ``anti-noise'' process introduced deliberately to the optomechanical force sensor. In the context of atomic spin measurements an analogous idea for coherent backaction cancellation was proposed independently~\cite{Julsgaard2001,Hammerer2009}, and applied for magnetometry below the standard quantum limit~\cite{Wasilewski2010}.

\begin{figure}[t]
	\subfigure[]{\includegraphics[width=0.7\columnwidth]{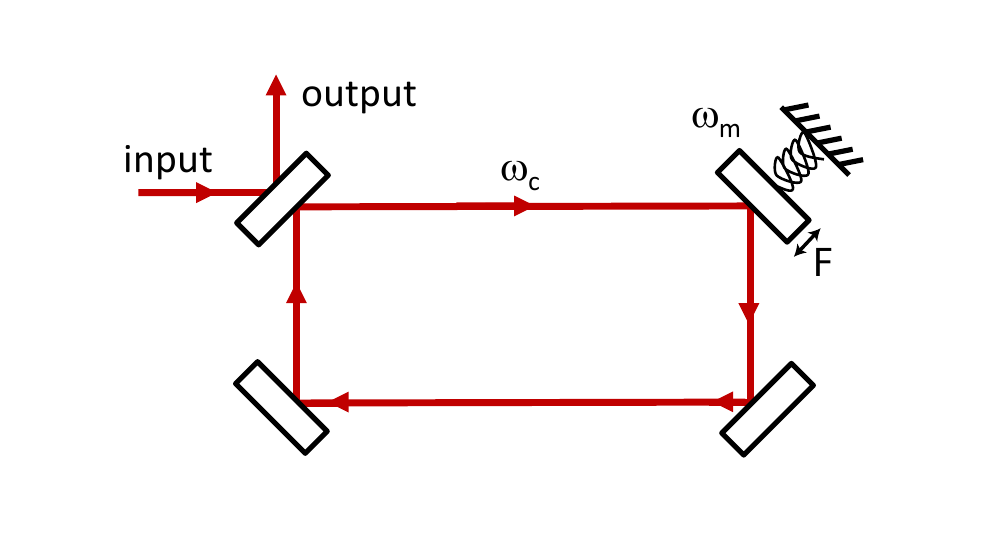}}\\
	\subfigure[]{\includegraphics[width=0.7\columnwidth]{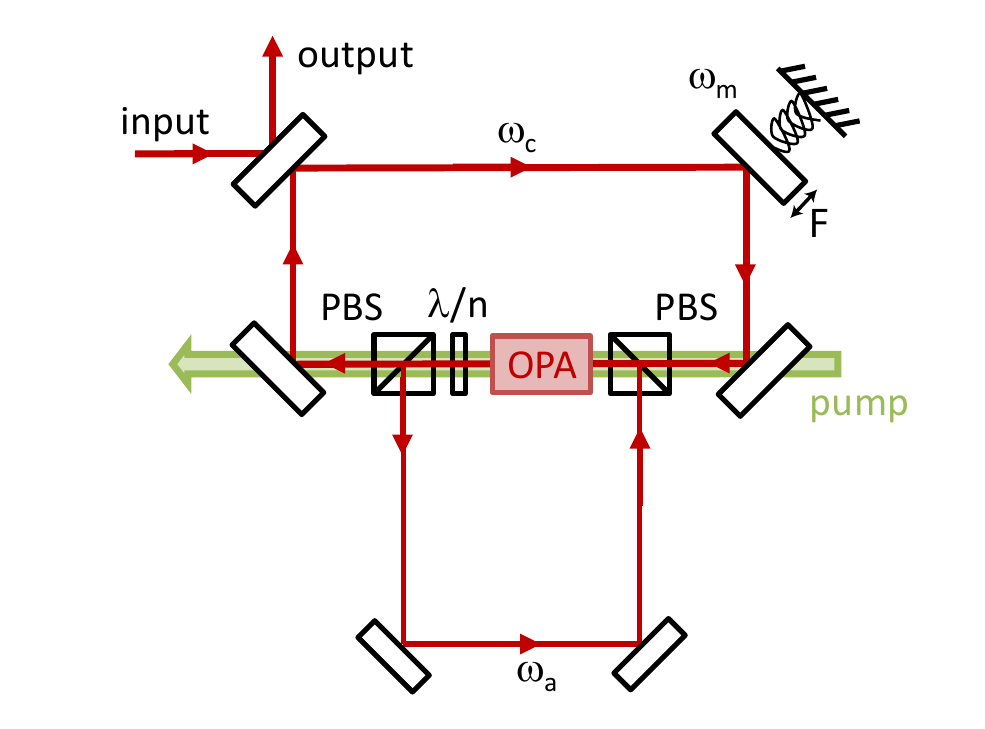}}\\
  \caption{(a)~Optomechanical cavity without noise cancellation. (b)~Cavity with coherent quantum noise cancellation. An auxiliary cavity coupled to the main cavity via a beamsplitter and an OPA-process provides an anti-noise path for backaction cancellation.}\label{fig:setup}
\end{figure}

In a conventional optomechanical detector force is estimated from its effect on the position on a test mass. The displacement of the test mass is read out optically from the phase shift it induces on light reflected off an optomechanical cavity, see Fig.~\ref{fig:setup}a. Backaction noise arises from amplitude (quantum) fluctuations of light which provide an additional (backaction) noise force on the test mass. For sufficiently intense probe light this force will be read out by light, too, and appear as a spurious signal (backaction noise) in the phase shift. The central idea in the scheme of Tsang \cite{Tsang2010} for backaction noise cancellation is to couple a second, auxiliary optical cavity to the optomechanical cavity which, too, is driven by amplitude fluctuations of probe light, but reacts to it in the exactly opposite way. This can be achieved if the ancillary system has a mass which is equal in magnitude to that of the test mass but is effectively \emph{negative}. The displacement of the ancillary system due to amplitude fluctuations will then be equal but opposite to that of the test mass, such that the two spurious phase shifts in the optomechanical cavity compensate each other exactly. Thus, backaction noise is coherently cancelled. In the scheme \cite{Tsang2010} the negative mass system is realized by means of an ancilla cavity coupled to the optomechanical cavity as shown schematically in Fig.~\ref{fig:setup}b and explained in more detailed below. Other setups which provide effective negative masses of ancilla systems for CQNC have been suggested and employ inverted atomic spins \cite{Hammerer2009} or Bose-Einstein condensates \cite{Zhang2013a}. On a more formal level, the mechanism can also be viewed as realizing an effectively classical dynamic in a subspace of an enlarged quantum system \cite{Tsang2012,Gough2009}.

In this article we consider in more detail the all-optical realization of CQNC by Tsang and Caves \cite{Tsang2010}, as shown in Fig.~\ref{fig:setup}b. For ideal CQNC the test mass and ancilla system should be equally but oppositely susceptible to amplitude fluctuations of light. Here we determine the experimental requirements of this matching of susceptibilities of the mechanical system and the cavity, and discuss the consequences of imperfect matching. Our main findings are as follows: While under ideal conditions the improvement below SQL can be on the order of the inverse mechanical quality factor of the test mass $Q_\m$, the improvement below SQL under realistic conditions is given by the ratio $\kappa_\a/\omega_\m$ of ancilla cavity line width to mechanical resonance frequency where the cavity line width is larger than the line width of the test mass resonance $\gamma_\m=\omega_\m/Q_\m<\kappa_\a$. Achieving a ratio $\kappa_\a/\omega_\m<1$ is possible with micromechanical oscillators, and we provide an experimental case study for a proof-of-principle demonstration of CQNC with such systems. However, for the free mass limit ($\omega_\m\rightarrow 0$) as relevant to gravitational wave detectors the requirement on the ancilla cavity line width appears to be prohibitive.

In Sec.~\ref{sec:force-sensing} we introduce our model for optomechanical force sensing, and compare the standard case (Fig.~\ref{fig:setup}a) subject to the SQL with the suggested setup (Fig.~\ref{fig:setup}b) for CQNC. We determine the central conditions for ideal CQNC on the physical parameters of the system. In Sec.~\ref{sec:force-sensing} we discuss the feasibility of these requirements, determine the impact of violations of the conditions for ideal CQNC, and provide am experimental case study for a micromechanical implementation.

\section{Force Sensing with and without backaction noise}\label{sec:force-sensing}

\subsection{Model}\label{sec:model}

We consider a generic optomechanical force detector as shown in Fig.~\ref{fig:setup}. The force $\mathrm{F}(t)$ to be detected acts on a harmonically bound probe mass $m$ whose position $X(t)$ and momentum $P(t)$ obey
\begin{align*}
  \dot{X}&=P/m\\
  \dot{P}&=-m\omega_m^2X-\gamma_\m P+\mathrm{f}_T+\mathrm{F}.
\end{align*}
$\omega_m$ is the oscillation frequency, and $\gamma_\m$ the damping rate of the mechanical oscillator. $\mathrm{f}_T(t)$ is the thermal Langevin force associated with the damping. For simplicity, we assume a white thermal noise force, such that $\langle \mathrm{f}_T(t)\mathrm{f}_T(t')\rangle=2\gamma_\m m k_BT\delta(t-t')$ (valid in the high temperature limit). In the following it will be convenient to work with dimensionless position and momentum variables $x_\m=X/x_\ZPF$ and $p_\m=P x_\ZPF/\hbar$ where the zero point fluctuation is $x_\ZPF=\sqrt{\hbar/m\omega_\m}$, such that $[x_\m,p_\m]=i$ and
\begin{equation}\label{eq:EOMsMecOsc}
\begin{split}
  \dot{x}_\m&=\omega_\m p_\m\\
  \dot{p}_\m&=-\omega_\m x_\m-\gamma_\m p_\m+\sqrt{\gamma_\m}(f_T+F).
\end{split}
\end{equation}
We have introduced scaled force operators $f_T=\mathrm{f}_T/\sqrt{\hbar m\gamma_\m\omega_\m}$ and $F=\mathrm{F}/\sqrt{\hbar m\gamma_\m\omega_\m}$ with dimension Hz$^{1/2}$. The scaled thermal force obeys $\langle {f}_T(t){f}_T(t')\rangle=\bar{n}\delta(t-t')$ where $\bar{n}=k_BT/\hbar\omega_\m$ is the mean number of phonons in thermal equilibrium \footnote{In the high temperature limit considered here $\bar{n}\gg1$.}.

The force $F(t)$ applied to the oscillator is estimated from a continuous weak measurement of the position of the probe mass accomplished through optical readout. We assume the oscillating test mass is one of the mirrors of an optical cavity, cf.~Fig.~\ref{fig:setup}a. The length change of the cavity due to the force $F(t)$ applied on the mirror is detected as a phase change in the light reflected off the cavity such that the force can be estimated from a phase sensitive homodyne detection of the light field. The precision of the corresponding estimate will be bounded by the standard quantum limit (SQL) due to the measurement backaction, as discovered by C. Caves \cite{Caves1980a}. We will summarize the quantitative formulation of the SQL as relevant to force detection as we go along. The setup for coherent backaction cancellation and sub-SQL force detection introduced by M.~Tsang and C.~Caves \cite{Tsang2010} is shown in Fig.~\ref{fig:setup}b. It  essentially corresponds to the standard setup of Fig.~\ref{fig:setup}a with an additional ancilla cavity properly coupled to the primary, optomechanical cavity used for position readout. In the following we will describe the dynamics of the system shown in Fig.~\ref{fig:setup}b, and concretize our results to the standard case without backaction cancellation shown in Fig.~\ref{fig:setup}a where appropriate.

The primary optomechanical cavity (also referred to as the meter cavity) will be described by dimensionless amplitude and phase quadratures $[x_\c,p_\c]=i$, and the ancilla cavity used for backaction cancellation by $[x_\a,p_\a]=i$. We will also use the corresponding annihilation operators $c=(x_\c+ip_\c)/\sqrt{2}$, and $a=(x_\a+ip_\a)/\sqrt{2}$. The dynamics of the two cavities, their mutual coupling and the coupling to the mechanical oscillator, are described by the Hamiltonian
\begin{align}\label{eq:Ham}
  H&=-\Delta a^\dagger a+g x_\c\, x_\m+g_\BS (ac^\dagger+a^\dagger c)
  +g_\DC(ac+a^\dagger c^\dagger).
\end{align}
which is written here in a frame rotating at the resonance frequency $\omega_\mathrm{c}$ of the meter cavity. $\Delta=\omega_\c-\omega_\a$ is the detuning of the meter cavity from the frequency $\omega_\a$ of the ancilla cavity. The term proportional to $g$ describes the radiation pressure interaction between the primary cavity and the mirror. Its strength is given by $g=\omega_\c x_\ZPF\alpha_\c/L$ where $L$ is the cavity length and $\alpha_\c$ the field amplitude in the meter cavity. The latter is given by $\alpha_\c=\sqrt{P/\hbar\omega_\c\kappa_c}$ where $P$ is the input power driving the cavity and $\kappa_\c$ is its line width. For a derivation of this standard treatment of the radiation pressure interaction in optomechanical systems we refer to \cite{Aspelmeyer2013}. The last two terms in the Hamiltonian \eqref{eq:Ham} describe the coupling of the ancilla cavity to the meter cavity as required for backaction cancellation and shown in Fig.~\ref{fig:setup}. The first term describes a passive beamsplitter-like mixing of the two cavity modes, while the second term denotes an active down conversion dynamic of the two modes through a non-degenerate optical parametric amplifier (OPA). The strength of the beamsplitter process is given by $g_\BS=rc/L$ with $r$ being the beamsplitter reflectivity and $c$ the speed of light. Without loss of generality we assume the length $L$ of ancilla cavity and meter cavity to be equal. The nonlinear coupling strength is $g_\DC=\Gamma lc/L$ with crystal length $l$ and gain parameter $\Gamma$ (see Sec.~\ref{sec:experiment}).

The Hamiltonian \eqref{eq:Ham} implies the equations of motion for the field quadratures of the meter and the ancilla cavity
\begin{align}\label{eq:linEOMs}
   \dot x_\c&=-\textstyle{\frac{\kappa_\c}{2}}  x_\c+(g_\BS-g_\DC) p_\a-\sqrt{\kappa_\c}x_\c^\in,\\
   \dot p_\c&=-\textstyle{\frac{\kappa_\c}{2}}  p_\c-(g_\BS+g_\DC) x_\a-g  x_\m-\sqrt{\kappa_\c}p_\c^\in,\label{eq:EOMpc}\\
   \dot x_\a&=-\textstyle{\frac{\kappa_\a}{2}}  x_\a+ \Delta p_\a+(g_\BS-g_\DC) p_\c-\sqrt{\kappa_\a}x_\a^\in,\\
   \dot p_\a&=-\textstyle{\frac{\kappa_\a}{2}}  p_\a- \Delta x_\a-(g_\BS+g_\DC) x_\c-\sqrt{\kappa_\a}p_\a^\in,\label{eq:EOMpa}
\intertext{and adds an additional term to the equation of motion \eqref{eq:EOMsMecOsc} of the mechanical oscillator}
   \dot x_\m&=\omega_\m  p_\m,\label{eq:EOMxm}\\
   \dot p_\m&=-\omega_\m  x_\m-\gamma_\m  p_\m+i[H,p_\m]+\sqrt{\gamma_\m}(f+F)\nonumber\\
   &=-\omega_\m  x_\m-\gamma_\m  p_\m-g  x_\c+\sqrt{\gamma_\m}(f+F),\label{eq:EOMpm}
\end{align}
In Eqs.~\eqref{eq:linEOMs} to \eqref{eq:EOMpa} we also added the decay of cavity modes at rates $\kappa_\a$ and $\kappa_\c$ along with white vacuum noise processes $a_\in(t)=(x_\a^\in+ip_\a^\in)/\sqrt{2}$ and $c_\in(t)=(x_\c^\in+ip_\c^\in)/\sqrt{2}$. The noise processes fulfill $\langle a_\in(t)a_\in^\dagger (t')\rangle=\delta(t-t')$, and $\langle c_\in(t)c_\in^\dagger (t')\rangle=\delta(t-t')$.

Eqs.~\eqref{eq:linEOMs} to \eqref{eq:EOMpm} describe the coupled dynamics of the mirror and the intracavity fields. The force $F(t)$ in Eq.~\eqref{eq:EOMpm} imprints, through Eq.~\eqref{eq:EOMxm}, its trace in the phase quadrature of the meter cavity $p_\c$, cf. Eq.~\eqref{eq:EOMpc}. It can therefore be detected in the output field of the meter cavity, which follows from the intracavity phase quadrature by means of the input-output relation \cite{Gardiner2004}
\begin{align}\label{eq:InOut}
 p_\c^\out(t)&=p_\c^\in(t)+\sqrt{\kappa_c}p_\c(t).
\end{align}
We note that the optical measurement of the force $F(t)$ comes at the cost of an additional force acting on the mirror proportional to the amplitude quadrature of the meter cavity $x_\c$, cf. Eq.~\eqref{eq:EOMpm}. This is the measurement backaction force ultimately giving rise to the standard quantum limit in force (or position) sensing.

The set of linear equations of motion \eqref{eq:linEOMs} to \eqref{eq:EOMpm} can be solved for the operators in frequency domain defined by
\begin{align*}
  p_\c(\omega)&=\frac{1}{\sqrt{2\pi}}\int\mathrm{d}t\,p_\c(t)e^{i\omega t},
\end{align*}
and analogously for all other quantities. In conjunction with the cavity input-output relation \eqref{eq:InOut} it is straightforward to determine the solution for the phase quadrature of the output field of the meter cavity $p_\c^\out(\omega)$ in its dependence on all the forces driving the system. Details of this calculation are given in Appendix \ref{Appendix}.

\subsection{Force Sensing without CQNC: The Standard Quantum Limit}

We consider first the standard setup for optomechanical force sensing without CQNC. This corresponds to a setup without ancilla cavity (Fig.~\ref{fig:setup}a) or, formally equivalent, with an ancilla cavity which is not coupled to the meter cavity,
\[
g_\BS=g_\DC=0.
\]
In this case the output phase quadrature in frequency space (see App.~\ref{Appendix}) is
\begin{align}
  p_\c^\out&=e^{i\phi}p_\c^\in
  +\sqrt{\gamma_\m \kappa_\c}g \chi_\c \chi_\m [f_T + F]
  +\kappa_\c g^2\chi_\c^2 \chi_\m x_\c^\in,\label{eq:pcoutomega}
\end{align}
where $e^{i\phi}=(i\omega-\frac{\kappa_c}{2})/(i\omega+\frac{\kappa_c}{2})$. We have defined the susceptibilities of the meter cavity $\chi_\c$ and the mechanical oscillator $\chi_\m$,
\begin{align}
  \chi_\c(\omega)&=\left[i\omega+\textstyle{\frac{\kappa_c}{2}}\right]^{-1},\label{eq:chic}\\
  \chi_\m(\omega)&=\omega_\m\left[(\omega^2-\omega_\m^2)-i\omega\gamma_\m\right]^{-1}.\label{eq:chim}
\end{align}
The frequency dependence of field quadratures, forces, and susceptibilities in Eq.~\eqref{eq:pcoutomega} is understood to be $\omega$ unless stated differently. Therefore we will drop the explicit mention of $\omega$ in the following equations. The four terms on the right hand side of Eq.~\eqref{eq:pcoutomega} describe, respectively, phase (shot) noise of light, thermal noise from Brownian motion of the mirror, the signal force to be detected, and backaction noise proportional to the amplitude quadrature of the input field.

In order to simplify the equations we will consider the limit where the linewidth of the meter cavity is larger than any measurement frequency of interest, $\kappa_\c\gg\omega$. In this case the susceptibility \eqref{eq:chic} becomes $\chi_\c\simeq 2/\kappa_c$ and it will be useful to introduce the effective measurement strength
\begin{align*}
  G=\frac{4g^2}{\kappa_\c}=\left(\frac{2}{\pi}\right)^2\frac{\omega_\c\mathcal{F}_\c P}{\omega_\m mc^2}
\end{align*}
where $\mathcal{F}_\c=\pi c/\kappa_c L_\c$ is the Finesse of the meter cavity. Expression \eqref{eq:pcoutomega} for the phase quadrature simplifies to
\begin{align}
  p_\c^\out&=e^{i\phi}p_\c^\in
  +\sqrt{\gamma_\m G}\chi_\m [f_T + F]
  +G \chi_\m x_\c^\in.\label{eq:pcoutomega1}
\end{align}
We emphasize that the assumption $\kappa_\c\gg\omega$ can be easily dropped at the cost of a somewhat more clumsy notation. The conclusions drawn in the following regarding achievable sensitivities are general and also hold also for measurement frequencies larger than the meter cavity line width.

Given the measurement of the phase quadrature $p_\c^\out(\omega)$ the optimal unbiased estimator $\hat{F}(\omega)$ of the force $F(\omega)$ is
\begin{align}\label{eq:optest}
\hat{F}&=\frac{1}{\sqrt{\gamma_\m G\phantom{|}}\chi_\m}p_\c^\out=F+F^\mathrm{add}
\end{align}
where the added force noise $F^\mathrm{add}(\omega)$ at detection frequency $\omega$ is
\begin{align*}
F^\mathrm{add}&=f+\frac{e^{i\phi}}{\sqrt{\gamma_\m G\phantom{|}}\chi_\m}p_\c^\in
+\sqrt{\frac{G}{\gamma_\m}} x_\c^\in
\end{align*}
Noise is added due to the thermal Langevin force (first term), shot noise in the phase quadrature (second term) and backaction from amplitude noise (third term).

\begin{figure}
  \includegraphics[width=0.9\columnwidth]{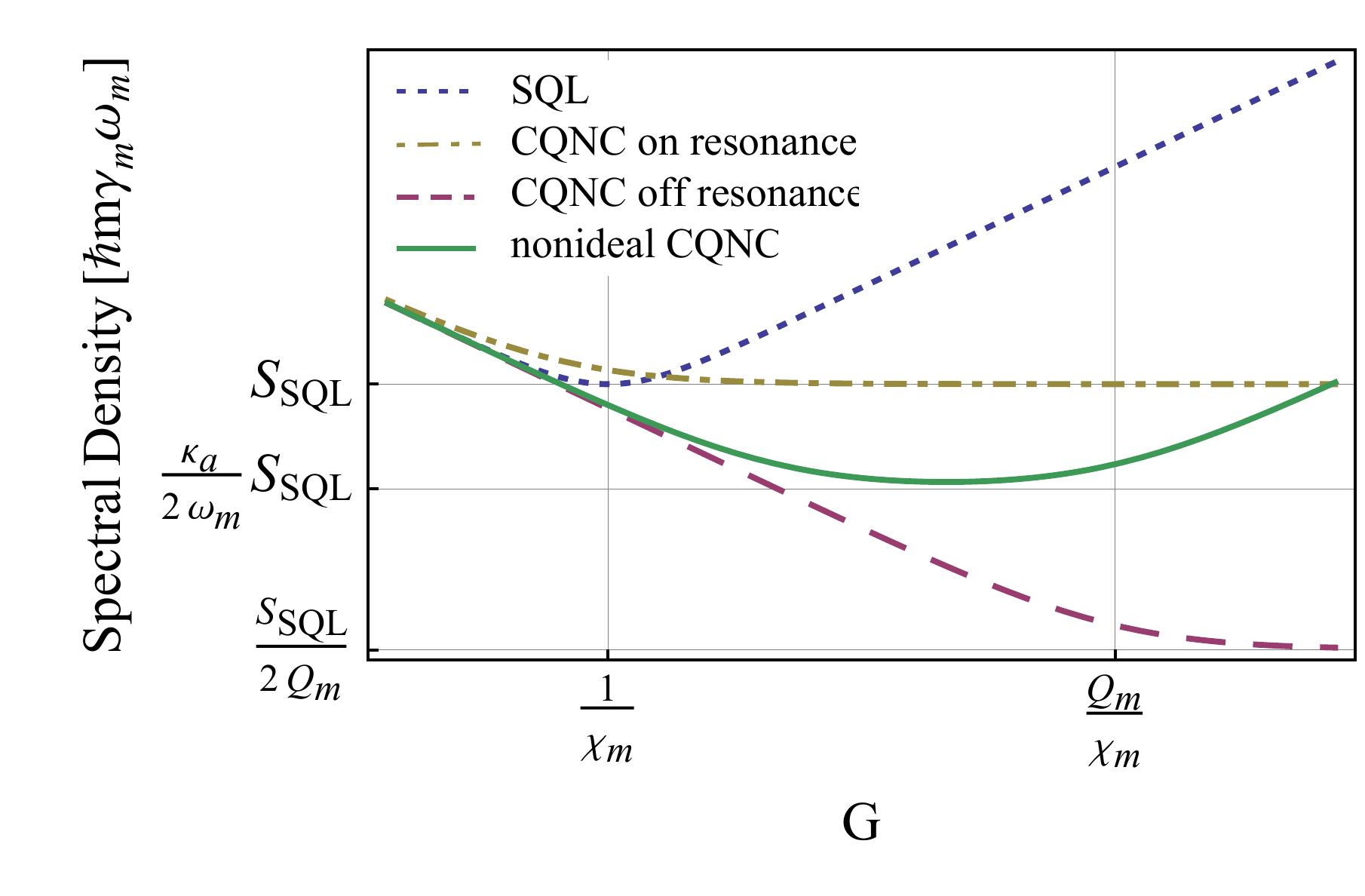}\\
  \caption{Noise power spectral densities $S_F(\omega)$ of force measurement versus measurement strength $G$ ($\propto$ power) at a fixed frequency $\omega$. The spectral density without coherent noise cancellation (Eq.~\eqref{eq:Sadd}, dotted blue line) exhibits its minimum at the standard quantum limit $S_\SQL=1/\gamma_\m|\chi_\m|$, Eq.~\eqref{eq:SSQL}, at an optimal $G=1/\chi_\m$. The spectral density with coherent noise cancellation, Eq.~\eqref{eq:SFcqnc}, is shown for $\omega=\omega_\m$ (yellow dot-dashed line) and for off-resonant frequencies (dashed purple). On resonance no improvement below the SQL is possible, off resonance the SQL can be surpassed by up to a factor of $1/2Q_\m$. In case of imperfect CQNC (solid green line) due to an ancilla cavity line width larger than the mechanical line width, $\kappa_\a>\gamma_\m$, the spectral density in Eq.~\ref{eq:SFcqncnonideal} will be limited by back action noise, but the minimum still falls below SQL if $\kappa_\a<2\omega_\m$. Here $\omega_\m/\kappa_\a=0.1$.} \label{Spectral-SQL-cqnc-kappa-limits_legend1}
\end{figure}

The sensitivity of the force measurement is commonly quantified by the (power) spectral density of added noise $S_F(\omega)$ defined through
\[
S_F(\omega)\delta(\omega-\omega')=\frac{1}{2}\langle F^\mathrm{add}(\omega)F^\mathrm{add}(-\omega')\rangle+\mathrm{c.c.}
\]
Assuming that amplitude and phase quadratures are uncorrelated and shot noise limited as given above the added noise spectral density is
\begin{align}\label{eq:Sadd}
  S_F&=\frac{k_BT}{\hbar\omega_\m}+\frac{1}{2\gamma_\m G|\chi_\m|^2}+\frac{G}{2\gamma_\m} .
\end{align}
As it stands, the noise spectral density is dimensionless. In order to convert this into a force noise spectral density $\mathrm{S}_F(\omega)$ in units of N$^2$Hz$^{-1}$ we have to multiply by the scale factor introduced in Sec.~\ref{sec:model}, such that $\mathrm{S}_F(\omega)=\hbar m\gamma_\m\omega_\m S_F(\omega)$ for the particular optomechanical force sensor. In view of Eq.~\eqref{eq:Sadd}, thermal Brownian noise provides a flat background to the force sensitivity independent of the measurement strength $G$, and therefore also independent of power, $G\propto P$.  Accordingly, measurement noise (due to phase noise) scales inversely proportional and measurement backaction noise (due to amplitude noise) scales proportionally to the power. The noise spectral density \eqref{eq:Sadd} is illustrated in Fig.~\eqref{Spectral-SQL-cqnc-kappa-limits_legend1}.

This implies that, for a given detection frequency $\omega$, there is an optimal value for the power used in the optical force readout. Minimizing the right hand side of Eq.~\eqref{eq:Sadd} with respect to power (or equivalently with respect to $G$) gives a lower bound for the achievable sensitivity,
\begin{align}\label{eq:SSQL}
  S_F(\omega)&\geq \frac{1}{\gamma_\m|\chi_\m(\omega)|}\equiv S_{\SQL}(\omega),
\end{align}
the \emph{standard quantum limit} (SQL) of continuous force sensing. Achieving this bound at a particular frequency $\omega$ requires a negligible thermal force (smaller than $S_{\SQL}(\omega)$) and a power such that the measurement strength is
\begin{align}\label{eq:GSQL}
G_{\SQL}(\omega)=\frac{1}{|\chi_\m(\omega)|}.
\end{align}

\subsection{Force Sensing with CQNC: Ideal Case}

Next, we want to consider the scheme for continuous force sensing with coherent backaction noise cancellation. In addition to the proposal of M. Tsang we consider the case of finite cavity linewidths and give limitations to the couplings and matchings of the optical setup. CQNC is ideally achieved if the ancilla cavity is coupled to the meter cavity in such a way that
\begin{align}\label{eq:ideal}
  g_\BS&=g_\DC & \mathrm{and}&& g&=g_\BS+g_\DC.
\end{align}
The rationale behind this particular choice will be made clear further below.
%
%
Assuming conditions \eqref{eq:ideal} the output phase quadrature is again easily found to be (cf. App.~\ref{Appendix})
\begin{equation}\label{eq:pcoutCQNC}
\begin{split}
  p_\c^\out&=e^{i\phi}p_\c^\in
  +\sqrt{\gamma_\m G} \chi_\m [f_T + F]\\
  &\quad-\sqrt{2\kappa_\a G}\chi_\a \left[\frac{i\omega+\kappa_\a/2}{\Delta} x^\in_\a+p^\in_\a\right]\\
  &\quad
  +G\left[\chi_\m+\chi_\a\right] x_\c^\in,
\end{split}
\end{equation}
which generalizes Eq.~\eqref{eq:pcoutomega}. Two new terms have appeared: The measurement backaction (last line) now contains a contribution which scales with the susceptibility of the ancilla cavity
\begin{align}\label{eq:chia}
\chi_\a(\omega)=\Delta\left[(\omega^2-\Delta^2-\textstyle{\frac{\kappa^2_\a}{4}})-i\omega\kappa_\a\right]^{-1}
\end{align}
defined here in analogy to the mechanical susceptibility $\chi_\m$ in Eq.~\eqref{eq:chim}. Overall the backaction now is proportional to the \emph{difference} of the mechanical susceptibility and that of the ancilla cavity. This means that measurement backaction affects the measured phase quadrature through two interfering channels. The key to the desired backaction cancellation will be to ensure this interference to be destructive. The other new contribution in Eq.~\eqref{eq:pcoutCQNC} (next to last term) corresponds to shot noise from the ancilla cavity transferred to the phase quadrature of the meter cavity, and will ultimately set the new limit to measurement sensitivity, as will be explained below.

The optimal estimator for the force $F(\omega)$ will still be given by Eq.~\eqref{eq:optest} with the phase quadrature now given by Eq.~\eqref{eq:pcoutCQNC}. Thus the added noise becomes
\begin{align}\label{eq:idealest}
F^\mathrm{add}&=f_T + \frac{e^{i\phi}}{\sqrt{\gamma_\m G\phantom{|}}\chi_\m}p_\c^\in\nonumber\\
&\quad -\sqrt{\frac{2\kappa_\a}{\gamma_\m}}\frac{\chi_\a}{\chi_\m}\left[\frac{i\omega+\kappa_\a/2}{\Delta} x^\in_\a+p^\in_\a\right]\nonumber\\
&\quad +\sqrt{\frac{G}{\gamma_\m}}\frac{\chi_\m+\chi_\a}{\chi_\m} x_\c^\in.
\end{align}
If conditions are such that
\begin{align}\label{eq:cond}
\chi_\m(\omega)=-\chi_\a(\omega)
\end{align}
for all $\omega$, the last term in Eq.~\eqref{eq:idealest} vanishes and backaction will be completely cancelled in the force measurement. In view of the explicit forms of the susceptibilities of the mechanical oscillator and the ancilla cavity, Eqs.~\eqref{eq:chim} and \eqref{eq:chia} respectively, condition \eqref{eq:cond} for ideal backaction noise cancellation occurs only if\\
(\emph{i}) the ancilla cavity is detuned from the meter cavity by
\begin{subequations}\label{eq:conds}
\begin{align}
  \Delta=-\omega_\m,
\end{align}
(\emph{ii}) the ancilla cavity linewidth matches that of the mechanical oscillator,
\begin{align}\label{eq:damping}
  \kappa_\a=\gamma_\m,
\end{align}
(\emph{iii}) and if $|\Delta|\gg\kappa_\a$. Due to (\emph{i}) and (\emph{ii}) this implies both, the resolved sideband limit
\begin{align}
\omega_\m&\gg\kappa_\a \label{eq:sideband-resolved},
\end{align}
and a large mechanical quality factor,
\begin{align}
Q_\m&=\omega_\m/\gamma_\m\gg 1.
\end{align}
\end{subequations}

\begin{figure}
  \includegraphics[width=0.9\columnwidth]{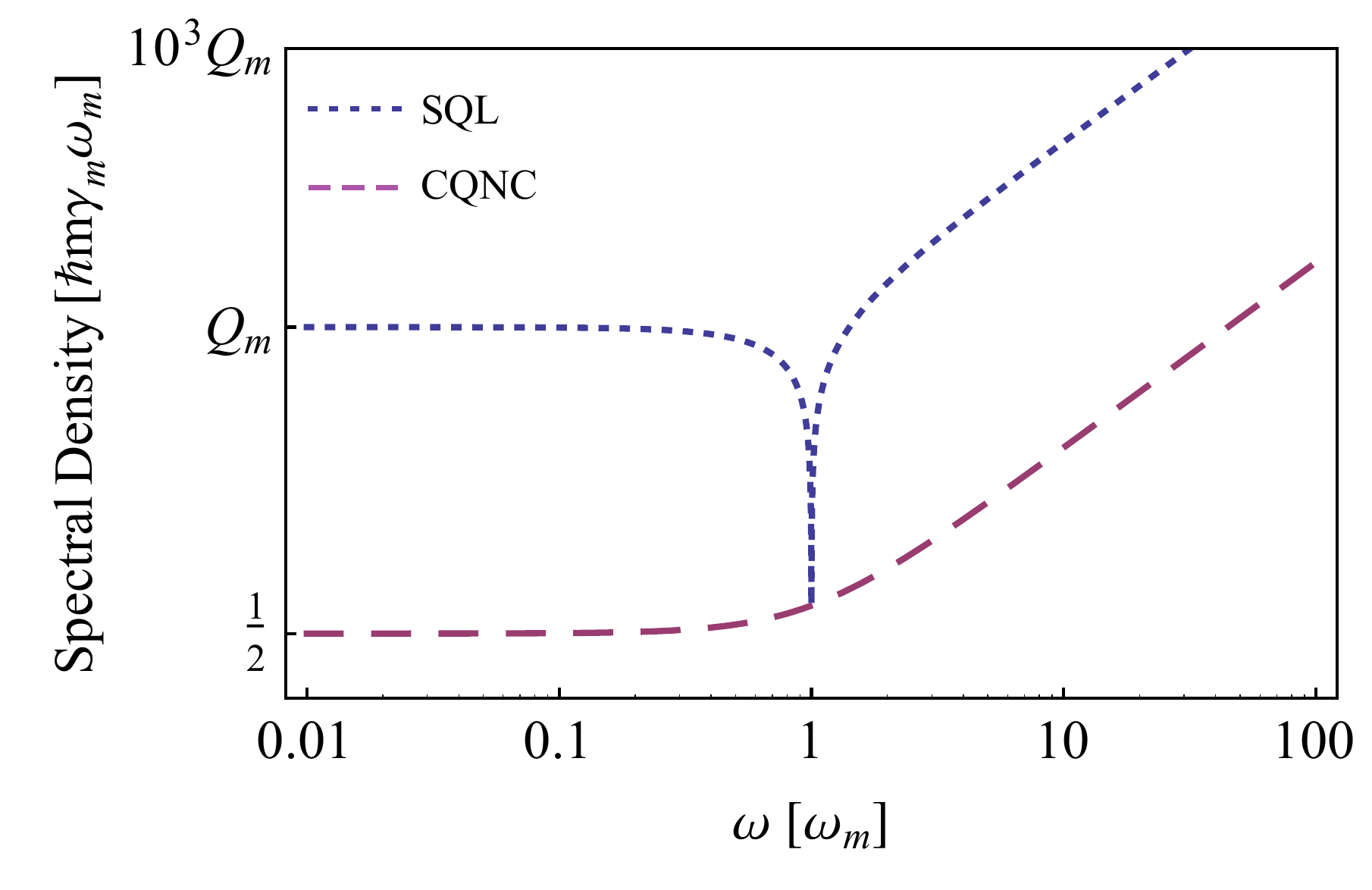}\\
  \caption{Noise power spectral densities achieved for optimal power versus frequency. The standard quantum limit $S_\SQL(\omega)$ (dotted blue, Eq.~\eqref{eq:SSQL}) can be surpassed with coherent quantum noise cancellation $S_\CQNC(\omega)$ (dashed purple, Eq.~\eqref{eq:SCQNC}) by a factor of $1/2Q_\m$ off resonance. Here $Q_\m=1000$.} \label{fig1}
\end{figure}

The feasibility of these conditions will be discussed in more detail below. While it is clear that not all of these conditions can be met perfectly we assume at least for the moment that conditions (\emph{i-iii}) hold and therefore \eqref{eq:cond} is fulfilled. Under this idealized assumption of perfect coherent quantum noise cancellation the last term in \eqref{eq:idealest} will not contribute. Assuming again that thermal noise is negligible the added noise consists only of contributions from the second and third term in \eqref{eq:idealest}, i.\,e. shot noise in the measured phase quadrature and shot noise introduced through the ancilla cavity,
\begin{align}\label{eq:SFcqnc}
S_F=\frac{1}{2\gamma_\m G|\chi_\m|^2}
+\frac{\kappa_\a|\chi_\a|^2}{\gamma_\m|\chi_\m|^2}\left[\frac{\omega^2+(\kappa_\a/2)^2}{\Delta^2}+1\right]
\end{align}
The shot noise contribution (first term) scales as $G^{-1}\propto P^{-1}$ and thus can be made negligible with respect to the second term for sufficiently large power. Taking into account conditions \eqref{eq:conds} we arrive at a lower bound for the sensitivity achievable with this method for coherent noise cancellation,
\begin{align}\label{eq:SCQNC}
S_F(\omega)&\geq \frac{\omega^2+\omega_\m^2+\gamma_\m^2/4}{\omega_\m^2}\equiv S_\CQNC(\omega).
\end{align}
The sensitivity $S_\CQNC(\omega)$ is compared to the standard quantum limit $S_\SQL(\omega)$ in Fig.~\ref{fig1}. From Eqs.~\eqref{eq:SSQL} and \eqref{eq:SCQNC} one finds in the parameter regime \eqref{eq:conds} considered here
\begin{align}\label{eq:Scqnc}
  S_\CQNC&=S_\SQL\times\begin{cases} \quad 1 & \text{on resonance}\quad \omega=\omega_\m \\ (2Q_\m)^{-1} & \text{off resonance}.\end{cases}
\end{align}
The enhancement in sensitivity can thus be on the order of the mechanical quality factor $Q_\m\gg1$ for frequencies away from the mechanical resonance. In order to make the second term in \eqref{eq:SFcqnc} larger than the measurement shot noise (first term) the measurement strength $G\propto P$ (power) has to fulfill
\begin{align}\label{eq:Gcqnc}
G_\CQNC&>\frac{\omega_\m^2}{2\gamma_\m|\chi_\m(\omega)|^2\left(\omega^2+\omega_\m^2+\gamma_\m^2/4\right)}\nonumber\\
&=G_\SQL\times\begin{cases} 1/2 & \omega=\omega_\m \\ Q_\m & \textrm{off resonance}.\end{cases}
\end{align}
We again used conditions \eqref{eq:conds} and the expression in Eq.~\eqref{eq:GSQL} for the measurement strength $G_\SQL$ necessary to achieve the standard quantum limit. As expected, the power level has to be increased by about the factor of the improvement in sensitivity. We emphasize that the sensitivity \eqref{eq:Scqnc} is achieved only asymptotically for power levels larger than the right hand side of Eq.~\eqref{eq:Gcqnc}. In particular, for the case on resonance a measurement strength of $G_\SQL$ will give a sensitivity which is actually \emph{worse} than the standard quantum limit, $S_F(\omega_\m)>S_\SQL(\omega_\m)$, as can be easily seen from Eq.~\eqref{eq:SFcqnc} and is also evident in Fig.~\ref{Spectral-SQL-cqnc-kappa-limits_legend1}. We conclude that the method of coherent noise cancellation considered here does not provide any advantage for resonant detections, but potentially provides a significant improvement for measurement at off-resonant frequencies. Accordingly we will restrict the subsequent discussion of imperfect noise cancellation to the case of measurement frequencies well above or below the mechanical resonance frequency.

\section{CQNC -- Requirements and Imperfections}\label{sec:requirements}
We now turn to a discussion of requirements and imperfections in the all-optical scheme for CQNC introduced in \cite{Tsang2010} and shown schematically in Fig.~\ref{fig:setup}b. Perfect CQNC requires the matching of coupling strengths of the meter cavity to the mechanical oscillator and the ancilla cavity, Eq.~\eqref{eq:ideal}, and matching of their susceptibilities, Eq.~\eqref{eq:cond} (or equivalently Eqs.~\eqref{eq:conds}). This raises the question what the tolerance is to violations of these conditions, and what the price is for a given mismatch. In particular condition \eqref{eq:damping} on the matching of the linewidth of the ancilla cavity to that of a high-quality mechanical oscillator appears to be very challenging, and can hardly be fulfilled without some compromise. After all, if the ultimate goal is to achieve backaction cancellation only within a certain frequency bandwidth, and not on the entire spectrum, it might in fact be advantageous to give up on one of the conditions and impose \eqref{eq:cond} only for the relevant frequencies.
%
%
\subsection{Nonideal ancilla cavity linewidth}\label{sec:nonidealancillacavitylinewidth}

\begin{figure}
\includegraphics[width=0.9\columnwidth]{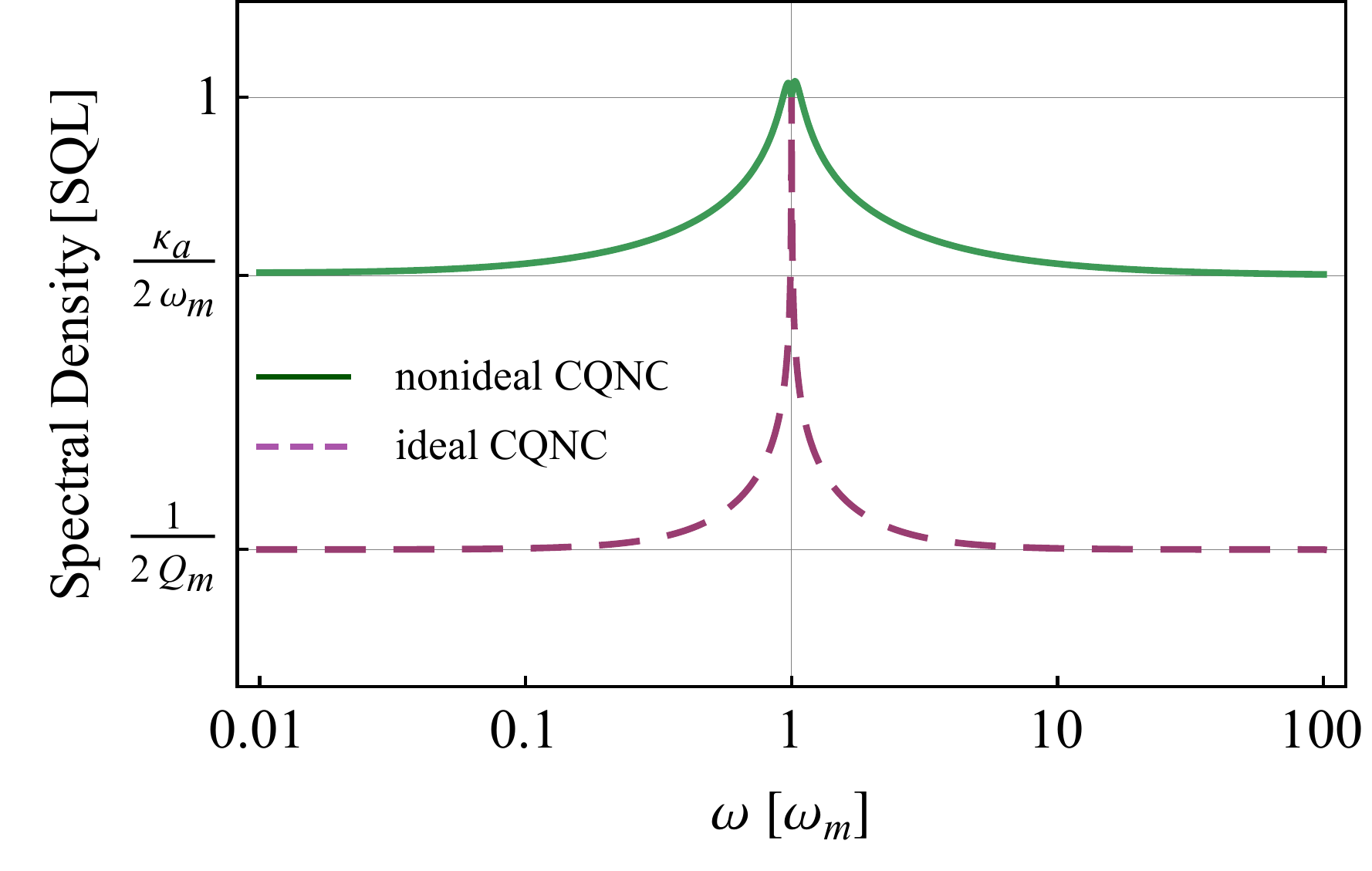}\\
  \caption{Spectral density of ideal (dashed purple, Eq.~\eqref{eq:SFcqnc}) and nonideal (solid green, \eqref{eq:SFcqncnonideal}) CQNC normalized to the standard quantum limit Eq.~\eqref{eq:SSQL}. On resonance no improvement can be achieved, whereas off resonance an improvement of $\kappa_\a/2\omega_\m$ is still possible.} \label{fig2}
\end{figure}

The hardest requirement to achieve in the all-optical setup for CQNC is the matching of the ancilla cavity linewidth to that of a mechanical oscillator. In particular in the regime of large mechanical quality factors typically employed for force sensing it is reasonable to expect $\kappa_\a\gg\gamma_\m$ instead of Eq.~\eqref{eq:damping}. Therefore, we will assume that condition \eqref{eq:cond} is violated due to a certain degree of mismatch in linewidths. The power spectral density of added noise corresponding to Eq.~\eqref{eq:idealest} then becomes
\begin{align}\label{eq:SFcqncnonideal}
  S_F&=\frac{1}{2\gamma_\m G|\chi_\m|^2}
+\frac{\kappa_\a|\chi_\a|^2}{\gamma_\m|\chi_\m|^2}\left[\frac{\omega^2+(\kappa_\a/2)^2}{\Delta^2}+1\right]\\
&\quad+\frac{G}{2\gamma_\m}\left|\frac{\chi_\m+\chi_\a}{\chi_\m}\right|^2,
\end{align}
where we neglect thermal noise. For ideal CQNC the trade-off with respect to the measurement strength $G$ vanishes with the last term in Eq.~\eqref{eq:pcoutCQNC} due to matching of the two susceptibilities. Imperfect backaction cancellation restores this trade-off which reduces the ability for noise reduction. As before one can derive a minimal spectral density achieved for optimal power,
\begin{align*}
  S_F&=\frac{|\chi_\m+\chi_\a|}{\gamma_\m|\chi_\m|^2}
+\frac{\kappa_\a|\chi_\a|^2}{\gamma_\m|\chi_\m|^2}\left[\frac{\omega^2+(\kappa_\a/2)^2}{\Delta^2}+1\right]
\end{align*}
For frequencies off resonance the second term (due to noise introduced by the ancilla cavity) dominates over the first term (measurement shot and backaction noise). Moreover, for an ancilla cavity linewidth larger than the mechanical frequency, $\kappa_\a\geq \omega_\m$, one can show that $S_F(\omega)\geq S_\SQL(\omega)$. In the opposite case, $\kappa_\a<\omega_\m$, an improvement in sensitivity is possible and one finds
\begin{align*}
S_F&=\frac{\kappa_\a}{2\omega_\m}\times S_\SQL
\end{align*}
(relevant for frequencies off resonance). This is illustrated in Fig.~\ref{fig2}. Thus, even for realistic linewidths of the ancilla cavity the improvement in sensitivity can be quite significant. A small ratio of cavity to mechanical linewidth $\kappa_\a/\omega_\m<1$ has been achieved for high-frequency mechanical oscillators in the context of the so-called resolved sideband limit of dynamic backaction cooling \cite{Aspelmeyer2013}. However, in the low frequency (free mass) limit this condition poses prohibitive requirements on the cavity Finesse and length.

\subsection{Imperfect matching of $g_{\BS}$ and $g_{\DC}$}
\begin{figure}
\includegraphics[width=0.9\columnwidth]{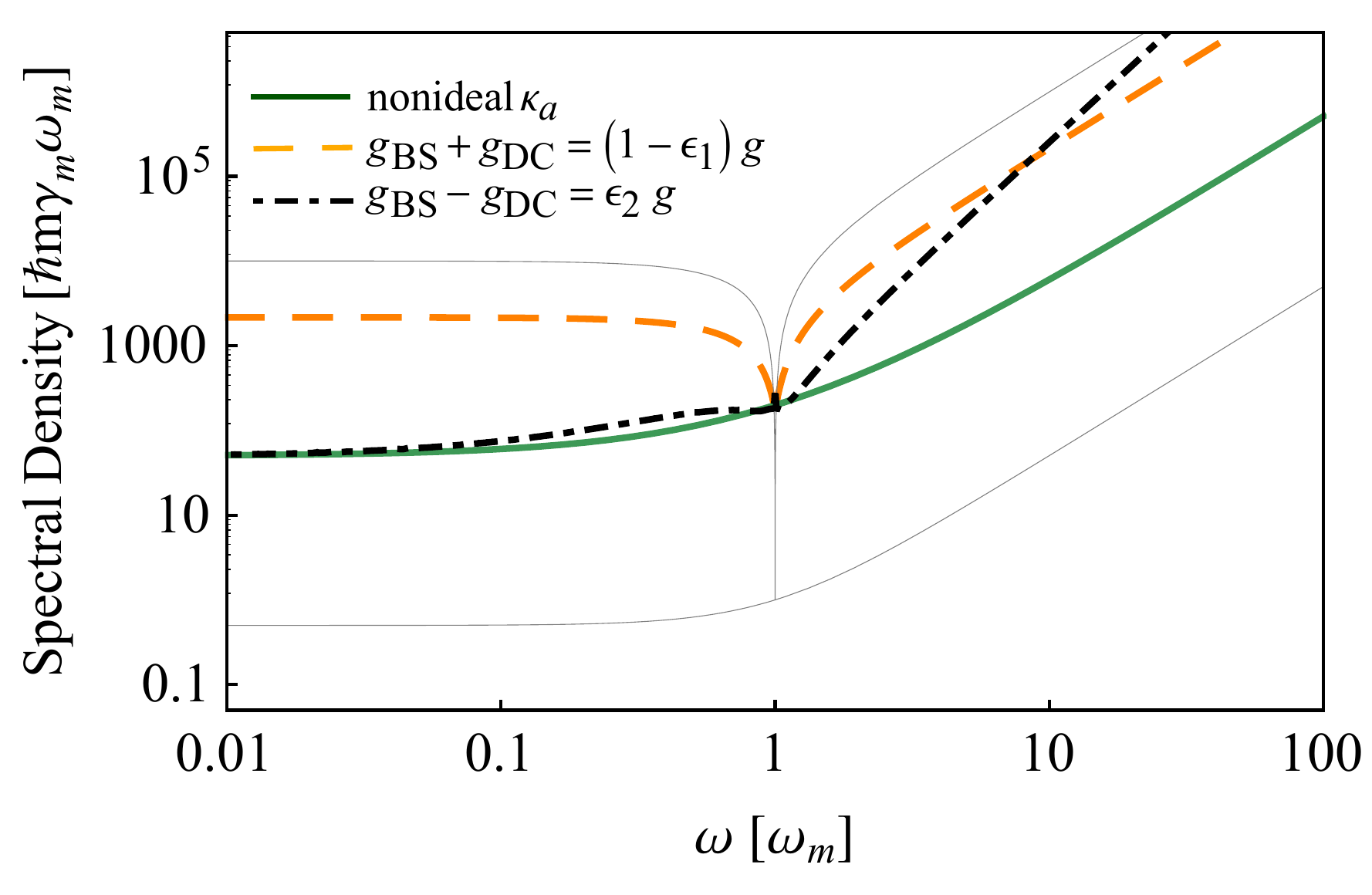}\\
  \caption{Noise spectral density for relative mismatch of $g_\BS$ and $g_\DC$ with respect to $g$ (dashed orange) and the relative mismatch between $g_\BS$ and $g_\DC$ (dot-dashed black). The mismatch $\epsilon_i$ is set to 10\,\% from the ideal value. As guidance SQL, ideal CQNC (both thin grey) and CQNC with nonideal $\kappa_\a$ (solid green) are plotted as well. The relative mismatch between $g_\BS$ and $g_\DC$ shows a freqency dependency which limits the sensitivity at higher frequencies. The mismatch to the optomechanical coupling gives a constant decrease of noise cancellation at all frequencies.}\label{fig3}
\end{figure}

The other important matching condition that needs to be fulfilled is twofold: The relative beamsplitter coupling $g_{\BS}$ must be matched with the down conversion coupling $g_{\DC}$, and the sum $g_\BS+g_\DC$ of the two must be matched with the optomechanical coupling $g$ to the oscillator, cf. Eqs.~\eqref{eq:ideal}. The optomechanical coupling strength $g=\sqrt{G\kappa_\c}/2\propto\sqrt{P}$ can be tuned with input power and the strength of the down conversion process $g_\DC$ depends on pump power, while the beam splitter coupling $g_\BS$ between meter and ancilla cavity is fixed for a given cavity geometry. Thus, matching of the three rates seems to be achievable. In order to determine the necessary level of precision we include a possible  mismatch in our calculations:
\begin{align}
g_\BS=\frac{1}{2}\left((1+\epsilon_1)g+\epsilon_2g\right)\\
g_\DC=\frac{1}{2}\left((1+\epsilon_1)g-\epsilon_2g\right).
\end{align}
In addition we assume a fixed finite ratio of $\kappa_\a/\omega_\m$, such that both conditions  Eq.~\eqref{eq:cond} and \eqref{eq:ideal} are violated. The resulting noise spectral density can be determined as explained in Appendix~\ref{Appendix} but the exact equations are involved and will not be reproduced here. The usual trade-off of measurement to backaction noise again requires optimization with respect to power. The dependence of the resulting minimal noise spectral density on a relative mismatch of 10\,\% for $1-\epsilon_1=(g_\BS+g_\DC)/g$ and $\epsilon_2=(g_\BS-g_\DC)/g$ is shown in Fig.~\ref{fig3} for a fixed ratio $\kappa_\a/\omega_\m=0.1$. While the mismatch for $1-\epsilon_1=(g_\BS+g_\DC)/g$ gives a constant decrease of noise cancellation for all frequencies the mismatch $\epsilon_2=(g_\BS-g_\DC)/g$ is frequency-dependent. For frequencies below the mechanical resonance a factor of $\kappa_\a/2\omega_\m$ can be achieved. At frequencies above the mechanical resonance the mismatch will be the limiting noise source for the measurement.

\subsection{Case Study}\label{sec:experiment}
After having evaluated ideal CQNC as well as the effect of a deviation from the ideal parameters in theory, we now discuss parameters for a proof-of-principle experiment showing the feasibility of CQNC. Recapitulating the requirements from the sections above for an experiment which realizes coherent quantum noise cancellation, ideally both the strengths of the coupling processes and the susceptibilities have to be matched,
\begin{equation}\label{eq:requirements}
g_\BS = g_\DC,\quad g_\BS+g_\DC=g\quad\text{and}\quad\chi_\m(\omega) = -\chi_\a(\omega),
\end{equation}
again with beamsplitter coupling $g_\BS=rc/L$, down conversion coupling $g_\DC=\Gamma l c/L$ and optomechanical coupling $g=\omega_\c x_\ZPF \alpha_\c/L$, where $c$ is the vacuum speed of light and $L$ the optical length of the main cavity. It is reasonable to assume that the reflectivity $r$ of $g_\BS$ is larger than some tenths of percent. This gives a lower bound for the other two coupling constants. The gain parameter $\Gamma$ \cite{Byer1977},
\begin{equation*}
\Gamma=\sqrt{\frac{2 \omega_1 \omega_2 d_\text{eff}^2 I_\text{pump}}{n_1 n_2 n_3 \epsilon_0 c^3}}
\end{equation*}
containing the refractive indices $n_i$ of signal, idler and pump beam, the frequencies $\omega_{1,2}$ of signal and idler beam and the nonlinear coefficient $d_\text{deff}$,
and thus $g_\DC$ can be tuned via the pump intensity $I_\text{pump}$ to match $g_\BS$. The sum of these two has to equal $g$, which depends on the oscillator's zero point fluctuation $x_\ZPF=\sqrt{\hbar/m\omega_\m}$ and can be balanced to a certain degree via the intracavity field amplitude $\alpha_\c=\sqrt{P/\hbar\omega_\c\kappa_\c}$.
The overall size of the coupling constants can additionally be adjusted via $L$ so that $g\leq\omega_\m$ due to stability considerations.

In view of the definition of the mechanical susceptibility and that of the ancilla cavity, Eqs.~\eqref{eq:chim} and \eqref{eq:chia}, conditions \eqref{eq:requirements} lead to the requirements
\begin{subequations}
\begin{align}
r c &= \Gamma l c = \frac12 \omega_\c x_\ZPF \alpha_\c\\
\Delta &= -\omega_\m\label{eq:matcheddetuning}
\end{align}
The prohibitive condition $\kappa_\a=\gamma_\m$ will be relaxed to
\begin{align}
\kappa_\a &< \omega_\m\label{eq:resolvedsideband}
\end{align}
\end{subequations}
as discussed in Sec.~\ref{sec:nonidealancillacavitylinewidth}.

\begin{table}
\caption{Proposed set of parameters.}
\setlength{\tabcolsep}{5pt}
\begin {tabular}{c l c c}
\toprule
$L$								& cavity length								& m					& $1.5$				\\
$r$ 							& beamsplitter reflectivity 	& \% 				& $0.5$				\\
$g_\BS/2\pi$ 			& beamsplitter coupling 			& kHz 			& $150$				\\
$I_\text{pump}$ 	& pump intensity 							& W/cm$^2$	& $45$				\\
$g_\DC/2\pi$ 			& down conversion coupling 		& kHz 			& $150$				\\
$P$ 							& cavity input power 					& mW 				& $100$				\\
$\kappa_\c/2\pi$ 	& meter cavity linewidth 			& MHz 			& $1$					\\
$\omega_\m/2\pi$	& mechanical resonance 				& MHz 			& $0.5$				\\
$\gamma_\m/2\pi$	& mechanical damping					& kHz				& $5$					\\
$m$ 							& effective mass 							& kg 				& $10^{-12}$	\\
$g/2\pi$ 					& optomechanical coupling 		& kHz 			& $300$				\\
$\kappa_\a/2\pi$	&	ancilla cavity linewidth		&	MHz				& $0.2$				\\
\hline
\end {tabular}
\label{tab:parameterlist}
\end{table}

For a first proof-of-principle experiment, the difficulties now lie on the one hand in identifying a suitable micromechanical resonator with a high zero point fluctuation, a high resonance frequency to realize the resolved sideband limit, and a large enough surface area ($\ge$50\,$\mu$m) as well as a high reflectivity to use it as an end-mirror in a cavity, on the other hand in reducing all the losses in the ancilla cavity to achieve the above mentioned resolved sideband limit. A small spotsize of about one third of the diameter of the optomechanical mirror on this mirror will be needed. This complicates using a long cavity which is necessary to achieve $g\le\omega_\m$. A challenging, yet feasible set of parameters is given in Tab.~\ref{tab:parameterlist}.

\begin{figure}
	\includegraphics[width=0.9\columnwidth]{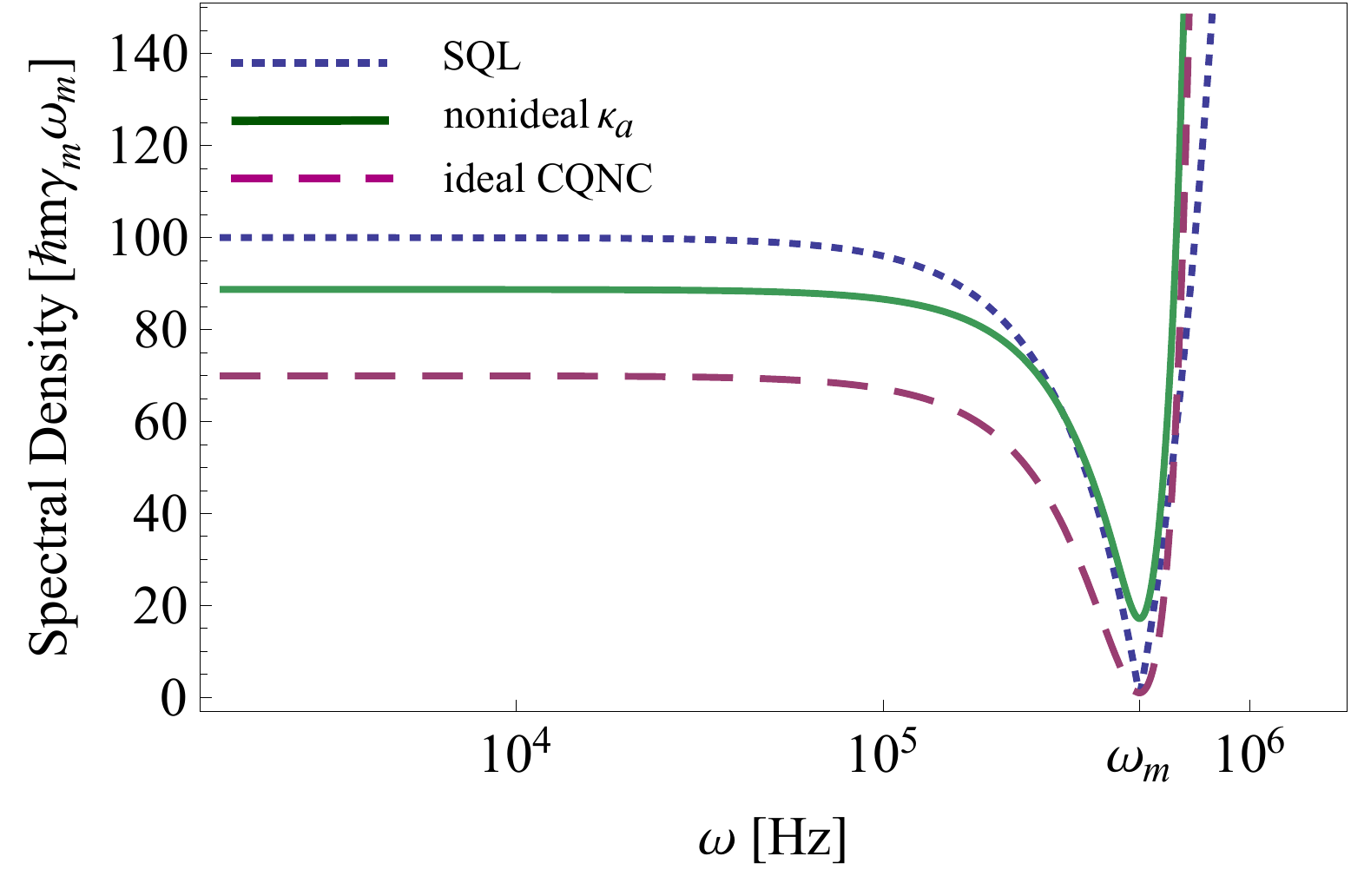}
  \caption{Quantum noise (dotted blue) and CQNC reduced noise for the system with parameters given in Table~\ref{tab:parameterlist}. Below 100\,MHz CQNC with nonideal $\kappa_\a$ (solid green) gives a radiation pressure backaction noise reduction of about 10\,\%. For ideal CQNC ($\kappa_\a$ matched, dashed purple) we achieve a noise reduction of 30\,\%. As we cannot increase the power to the optimal value the reduction is less than $\kappa_\a/2\omega_\m$ or $1/2Q_\m$ respectively.} \label{realnoise}
\end{figure}

For this set of parameters we theoretically achieve quantum noise reduction of 10\,\% at frequencies below the mechanical resonance of our oscillator (see Fig.~\ref{realnoise}). The system will be shot noise limited at frequencies above the mechanical resonance; here CQNC will have no effect on the sensitivity. To apply CQNC at higher frequencies or to achieve the calculated noise reduction of $\kappa_\a/\omega_\m$ we need to increase the power such that the system is radiation pressure backaction noise limited above the resonance frequency. However at this stage the optomechanical coupling will have reached a critical value which changes the optomechanical parameters of our system to unwanted bound conditions, like an optical spring.

To show the feasibility of CQNC we will need to set up a system which is radiation pressure limited -- a challenging requirement. To show that the noise cancellation scheme works in principle, we will instead use artificial amplitude noise increasing the backaction noise above the thermal bath. Assuming an input power of 100\,mW to our cavity, thermal noise at room temperature masks the SQL by five orders of magnitude.  To test our CQNC scheme we have to modulate our input beam with an amplitude of 100\,$\mu$W to increase the amplitude noise above the thermal noise. Generating this amount of noise is possible with common amplitude modulators.



\section{Conclusion}
Our calculations show that the experimental realization of the proposed feed-forward coherent noise cancellation scheme is feasible with existing technology. We have given a quantitative measure for the noise reduction which is achievable with ideal CQNC and showed that a noise reduction is still possible even if the requirements are not met perfectly. We have also given a set of parameters for a possible experiment which shows quantum noise reduction at frequencies below the mechanical resonance of the chosen optomechanical oscillator.\\
This research was supported by the Deutsche Forschungsgemeinschaft through the Centre for Quantum Engineering and Space-Time Research (QUEST), the Sonderforschungsbereich Transregio 7 ``Gravitational Wave Astronomy'' and by the Vienna Science and Technology Fund (WWTF) through project ICT12-049. The authors would like to acknowledge and wish to thank Elanor Huntington, Mankei Tsang, and Eugene Polzik for many helpful discussions and insightful comments.

%
%
\begin{appendix}
  \section{}\label{Appendix}
\subsection*{Noise Spectral Densities}
With the matrix $M$ of the system of differential equations and the noise matrix $A$,
\begin{widetext}
\begin{align*}
	M=\begin{pmatrix}
		-\kappa_\c/2 &	0	&	0	&	g_\BS-g_\DC	&	0	&	0\\
		0 &	-\kappa_\c/2 & -g_\BS-g_\DC & 0 & -g & 0\\
		0 & g_\BS-g_\DC & -\kappa_\a/2 & \Delta & 0 & 0\\
		-g_\BS-g_\DC & 0 & -\Delta & -\kappa_\a/2 & 0 & 0\\
		0 & 0 & 0 & 0 & 0 & \omega_\m\\
		-g & 0 & 0 & 0 & -\omega_\m & -\gamma_\m
		\end{pmatrix},\quad
	A=\begin{pmatrix}
		-\sqrt\kappa_\c &	0	&	0	&	0	&	0	&	0\\
		0 &	-\sqrt\kappa_\c & 0 & 0 & 0 & 0\\
		0 & 0 & -\sqrt\kappa_\a & 0 & 0 & 0\\
		0 & 0 & 0 & -\sqrt\kappa_\a & 0 & 0\\
		0 & 0 & 0 & 0 & 0 & 0\\
		0 & 0 & 0 & 0 & 0 & \sqrt{\gamma_\m}
	\end{pmatrix},
\end{align*}
\end{widetext}
the cavity quadratures $\vec x$, input field quadratures $\vec x_\in$ and output field quadratures $\vec x_\out$ being
\begin{equation*}
	\vec x=\begin{pmatrix}
		x_\c \\ p_\c \\ x_\a \\ p_\a \\ x_\m \\ p_\m
	\end{pmatrix},
	\quad
	\vec x_\in=\begin{pmatrix}
		x_\c^\in \\ p_\c^\in \\ x_\a^\in \\ p_\a^\in \\ 0 \\ f+F
	\end{pmatrix},
	\quad
	\vec x_\out=\begin{pmatrix}
		x_\c^\out \\ p_\c^\out \\ x_\a^\out \\ p_\a^\out \\ x_\m^\out \\ p_\m^\out
	\end{pmatrix}
\end{equation*}
the linearized system (\ref{eq:linEOMs}--\ref{eq:EOMpm}) can now be described as
\begin{equation}\label{eq:EOMmatrix}
\dot{\vec{x}}=M\vec x+A\vec x_\in.
\end{equation}
With a phase-sensitive measurement of the output beam, the quadratures $x_\c^\out$ and $p_\c^\out$ can be accessed. The force $F$ to be detected couples into momentum $p_\m$ of the mirror, and via $x_\m$ and $p_\c$ can now be found in the output phase quadrature $p_\c^\out$.

The input-output formalism,
\begin{equation}\label{eq:input-output}
\vec x_\out = \vec x_\in - A\vec x,
\end{equation}
together with a Fourier transformation of \eqref{eq:EOMmatrix} into the frequency domain, defined by
\begin{equation*}
\vec x(\omega) = \frac{1}{\sqrt{2\pi}} \int{\!\rm{d}t}\,\vec{x}(t) e^{i\omega t}
\end{equation*}
so that
\begin{equation*}
\dot{\vec{x}}=i\omega\vec x=M\vec x+A\vec x_\in,
\end{equation*}
leads to
\begin{align}\label{eq:solution}
\vec x_\out&=P\vec x_\in
\end{align}
where
\begin{align*}
P&=\mathbbm{1}-A\frac{1}{i\omega-M}A.
\end{align*}
The spectral density is given as $S_\out=PS_\in P^{T}$ where $S_\in=\frac{1}{2}\mathrm{diag}(1,1,1,1,0,2\bar{n})$. The phase spectral density $S_{pp}$ of the system output, which contains information about the force $F$, is defined by $S_{pp}(\omega)\delta(\omega-\omega^\prime)=\langle p_\c(\omega)p_\c(\omega^\prime)\rangle$ and represented by the (2\,,\,2)-element of $S_\out$.

In the special case $g_\DC=g_\BS=0$, the system reduces to (cf. Eqs.~(\ref{eq:linEOMs}--\ref{eq:EOMpm}))
\begin{align}
   \dot x_\c&=-\textstyle{\frac{\kappa_\c}{2}}  x_\c-\sqrt{\kappa_\c}x_\c^\in,	\label{eq:xc}\\
   \dot p_\c&=-\textstyle{\frac{\kappa_\c}{2}}  p_\c-g  x_\m-\sqrt{\kappa_\c}p_\c^\in,	\label{eq:pc}\\
   \dot x_\m&=\omega_\m  p_\m,	\label{eq:xm}\\
   \dot p_\m&=-\omega_\m  x_\m-\gamma_\m  p_\m-g  x_\c+\sqrt{\gamma_\m}(f+F).	\label{eq:pm}
\end{align}
Solving this system in frequency space by inserting \eqref{eq:xc} into \eqref{eq:pm}, \eqref{eq:pm} into \eqref{eq:xm}, and \eqref{eq:xm} into \eqref{eq:pc}, together with the input-output relation $p_\c^\out=p_\c^\in +\sqrt{\kappa_\c}p_\c$ leads to \eqref{eq:pcoutomega}. An analogous calculation can be performed for perfect CQNC, i.\,e. $g_\BS=g_\DC=\frac12 g$, to find \eqref{eq:pcoutCQNC}. For imperfect CQNC, the system cannot be reduced to a simple expression. We therefore calculate the exact solution computationally to generate our graphs for the imperfect case.
\end{appendix}

\bibliography{CQNC}

\begin{thebibliography}{35}%
\makeatletter
\providecommand \@ifxundefined [1]{%
 \@ifx{#1\undefined}
}%
\providecommand \@ifnum [1]{%
 \ifnum #1\expandafter \@firstoftwo
 \else \expandafter \@secondoftwo
 \fi
}%
\providecommand \@ifx [1]{%
 \ifx #1\expandafter \@firstoftwo
 \else \expandafter \@secondoftwo
 \fi
}%
\providecommand \natexlab [1]{#1}%
\providecommand \enquote  [1]{``#1''}%
\providecommand \bibnamefont  [1]{#1}%
\providecommand \bibfnamefont [1]{#1}%
\providecommand \citenamefont [1]{#1}%
\providecommand \href@noop [0]{\@secondoftwo}%
\providecommand \href [0]{\begingroup \@sanitize@url \@href}%
\providecommand \@href[1]{\@@startlink{#1}\@@href}%
\providecommand \@@href[1]{\endgroup#1\@@endlink}%
\providecommand \@sanitize@url [0]{\catcode `\\12\catcode `\$12\catcode
  `\&12\catcode `\#12\catcode `\^12\catcode `\_12\catcode `\%12\relax}%
\providecommand \@@startlink[1]{}%
\providecommand \@@endlink[0]{}%
\providecommand \url  [0]{\begingroup\@sanitize@url \@url }%
\providecommand \@url [1]{\endgroup\@href {#1}{\urlprefix }}%
\providecommand \urlprefix  [0]{URL }%
\providecommand \Eprint [0]{\href }%
\providecommand \doibase [0]{http://dx.doi.org/}%
\providecommand \selectlanguage [0]{\@gobble}%
\providecommand \bibinfo  [0]{\@secondoftwo}%
\providecommand \bibfield  [0]{\@secondoftwo}%
\providecommand \translation [1]{[#1]}%
\providecommand \BibitemOpen [0]{}%
\providecommand \bibitemStop [0]{}%
\providecommand \bibitemNoStop [0]{.\EOS\space}%
\providecommand \EOS [0]{\spacefactor3000\relax}%
\providecommand \BibitemShut  [1]{\csname bibitem#1\endcsname}%
\let\auto@bib@innerbib\@empty
\bibitem [{\citenamefont {Caves}\ \emph {et~al.}(1980)\citenamefont {Caves},
  \citenamefont {Thorne}, \citenamefont {Drever}, \citenamefont {Sandberg},\
  and\ \citenamefont {Zimmermann}}]{Caves1980a}%
  \BibitemOpen
  \bibfield  {author} {\bibinfo {author} {\bibfnamefont {C.}~\bibnamefont
  {Caves}}, \bibinfo {author} {\bibfnamefont {K.}~\bibnamefont {Thorne}},
  \bibinfo {author} {\bibfnamefont {R.}~\bibnamefont {Drever}}, \bibinfo
  {author} {\bibfnamefont {V.}~\bibnamefont {Sandberg}}, \ and\ \bibinfo
  {author} {\bibfnamefont {M.}~\bibnamefont {Zimmermann}},\ }\href
  {http://link.aps.org/doi/10.1103/RevModPhys.52.341} {\bibfield  {journal}
  {\bibinfo  {journal} {Reviews of Modern Physics}\ }\textbf {\bibinfo {volume}
  {52}},\ \bibinfo {pages} {341} (\bibinfo {year} {1980})}\BibitemShut
  {NoStop}%
\bibitem [{\citenamefont {Clerk}\ \emph {et~al.}(2010)\citenamefont {Clerk},
  \citenamefont {Devoret}, \citenamefont {Girvin}, \citenamefont {Marquardt},\
  and\ \citenamefont {Schoelkopf}}]{Clerk2010a}%
  \BibitemOpen
  \bibfield  {author} {\bibinfo {author} {\bibfnamefont {A.~A.}\ \bibnamefont
  {Clerk}}, \bibinfo {author} {\bibfnamefont {M.~H.}\ \bibnamefont {Devoret}},
  \bibinfo {author} {\bibfnamefont {S.~M.}\ \bibnamefont {Girvin}}, \bibinfo
  {author} {\bibfnamefont {F.}~\bibnamefont {Marquardt}}, \ and\ \bibinfo
  {author} {\bibfnamefont {R.~J.}\ \bibnamefont {Schoelkopf}},\ }\href
  {http://link.aps.org/doi/10.1103/RevModPhys.82.1155} {\bibfield  {journal}
  {\bibinfo  {journal} {Reviews of Modern Physics}\ }\textbf {\bibinfo {volume}
  {82}},\ \bibinfo {pages} {1155} (\bibinfo {year} {2010})}\BibitemShut
  {NoStop}%
\bibitem [{\citenamefont {Danilishin}\ and\ \citenamefont
  {Khalili}(2012)}]{Danilishin2012}%
  \BibitemOpen
  \bibfield  {author} {\bibinfo {author} {\bibfnamefont {S.~L.}\ \bibnamefont
  {Danilishin}}\ and\ \bibinfo {author} {\bibfnamefont {F.~Y.}\ \bibnamefont
  {Khalili}},\ }\href
  {http://relativity.livingreviews.org/Articles/lrr-2012-5/} {\bibfield
  {journal} {\bibinfo  {journal} {Living Rev. Relativity}\ }\textbf {\bibinfo
  {volume} {15}} (\bibinfo {year} {2012})}\BibitemShut {NoStop}%
\bibitem [{\citenamefont {Chen}(2013)}]{Chen2013}%
  \BibitemOpen
  \bibfield  {author} {\bibinfo {author} {\bibfnamefont {Y.}~\bibnamefont
  {Chen}},\ }\href {\doibase 10.1088/0953-4075/46/10/104001} {\bibfield
  {journal} {\bibinfo  {journal} {Journal of Physics B: Atomic, Molecular and
  Optical Physics}\ }\textbf {\bibinfo {volume} {46}},\ \bibinfo {pages}
  {104001} (\bibinfo {year} {2013})}\BibitemShut {NoStop}%
\bibitem [{\citenamefont {Aspelmeyer}\ \emph {et~al.}(2013)\citenamefont
  {Aspelmeyer}, \citenamefont {Kippenberg},\ and\ \citenamefont
  {Marquardt}}]{Aspelmeyer2013}%
  \BibitemOpen
  \bibfield  {author} {\bibinfo {author} {\bibfnamefont {M.}~\bibnamefont
  {Aspelmeyer}}, \bibinfo {author} {\bibfnamefont {T.~J.}\ \bibnamefont
  {Kippenberg}}, \ and\ \bibinfo {author} {\bibfnamefont {F.}~\bibnamefont
  {Marquardt}},\ }\href {http://arxiv.org/abs/1303.0733} {\ ,\ \bibinfo {pages}
  {65} (\bibinfo {year} {2013})},\ \Eprint {http://arxiv.org/abs/1303.0733}
  {arXiv:1303.0733} \BibitemShut {NoStop}%
\bibitem [{\citenamefont {Caves}(1981)}]{Caves1981}%
  \BibitemOpen
  \bibfield  {author} {\bibinfo {author} {\bibfnamefont {C.}~\bibnamefont
  {Caves}},\ }\href {\doibase 10.1103/PhysRevD.23.1693} {\bibfield  {journal}
  {\bibinfo  {journal} {Physical Review D}\ }\textbf {\bibinfo {volume} {23}},\
  \bibinfo {pages} {1693} (\bibinfo {year} {1981})}\BibitemShut {NoStop}%
\bibitem [{\citenamefont {Abramovici}\ \emph {et~al.}(1992)\citenamefont
  {Abramovici}, \citenamefont {Althouse}, \citenamefont {Drever}, \citenamefont
  {G\"{u}rsel}, \citenamefont {Kawamura}, \citenamefont {Raab}, \citenamefont
  {Shoemaker}, \citenamefont {Sievers}, \citenamefont {Spero}, \citenamefont
  {Thorne}, \citenamefont {Vogt}, \citenamefont {Weiss}, \citenamefont
  {Whitcomb},\ and\ \citenamefont {Zucker}}]{Abramovici1992}%
  \BibitemOpen
  \bibfield  {author} {\bibinfo {author} {\bibfnamefont {A.}~\bibnamefont
  {Abramovici}}, \bibinfo {author} {\bibfnamefont {W.~E.}\ \bibnamefont
  {Althouse}}, \bibinfo {author} {\bibfnamefont {R.~W.~P.}\ \bibnamefont
  {Drever}}, \bibinfo {author} {\bibfnamefont {Y.}~\bibnamefont {G\"{u}rsel}},
  \bibinfo {author} {\bibfnamefont {S.}~\bibnamefont {Kawamura}}, \bibinfo
  {author} {\bibfnamefont {F.~J.}\ \bibnamefont {Raab}}, \bibinfo {author}
  {\bibfnamefont {D.}~\bibnamefont {Shoemaker}}, \bibinfo {author}
  {\bibfnamefont {L.}~\bibnamefont {Sievers}}, \bibinfo {author} {\bibfnamefont
  {R.~E.}\ \bibnamefont {Spero}}, \bibinfo {author} {\bibfnamefont {K.~S.}\
  \bibnamefont {Thorne}}, \bibinfo {author} {\bibfnamefont {R.~E.}\
  \bibnamefont {Vogt}}, \bibinfo {author} {\bibfnamefont {R.}~\bibnamefont
  {Weiss}}, \bibinfo {author} {\bibfnamefont {S.~E.}\ \bibnamefont {Whitcomb}},
  \ and\ \bibinfo {author} {\bibfnamefont {M.~E.}\ \bibnamefont {Zucker}},\
  }\href {http://www.sciencemag.org/content/256/5055/325} {\bibfield  {journal}
  {\bibinfo  {journal} {Science}\ }\textbf {\bibinfo {volume} {256}},\ \bibinfo
  {pages} {325} (\bibinfo {year} {1992})}\BibitemShut {NoStop}%
\bibitem [{\citenamefont {Purdy}\ \emph {et~al.}(2013)\citenamefont {Purdy},
  \citenamefont {Peterson},\ and\ \citenamefont {Regal}}]{Purdy2013}%
  \BibitemOpen
  \bibfield  {author} {\bibinfo {author} {\bibfnamefont {T.~P.}\ \bibnamefont
  {Purdy}}, \bibinfo {author} {\bibfnamefont {R.~W.}\ \bibnamefont {Peterson}},
  \ and\ \bibinfo {author} {\bibfnamefont {C.~A.}\ \bibnamefont {Regal}},\
  }\href {\doibase 10.1126/science.1231282} {\bibfield  {journal} {\bibinfo
  {journal} {Science (New York, N.Y.)}\ }\textbf {\bibinfo {volume} {339}},\
  \bibinfo {pages} {801} (\bibinfo {year} {2013})}\BibitemShut {NoStop}%
\bibitem [{\citenamefont {Murch}\ \emph {et~al.}(2008)\citenamefont {Murch},
  \citenamefont {Moore}, \citenamefont {Gupta},\ and\ \citenamefont
  {Stamper-Kurn}}]{Murch2008}%
  \BibitemOpen
  \bibfield  {author} {\bibinfo {author} {\bibfnamefont {K.~W.}\ \bibnamefont
  {Murch}}, \bibinfo {author} {\bibfnamefont {K.~L.}\ \bibnamefont {Moore}},
  \bibinfo {author} {\bibfnamefont {S.}~\bibnamefont {Gupta}}, \ and\ \bibinfo
  {author} {\bibfnamefont {D.~M.}\ \bibnamefont {Stamper-Kurn}},\ }\href
  {\doibase 10.1038/nphys965} {\bibfield  {journal} {\bibinfo  {journal}
  {Nature Physics}\ }\textbf {\bibinfo {volume} {4}},\ \bibinfo {pages} {561}
  (\bibinfo {year} {2008})}\BibitemShut {NoStop}%
\bibitem [{\citenamefont {Harry}(2010)}]{Harry2010}%
  \BibitemOpen
  \bibfield  {author} {\bibinfo {author} {\bibfnamefont {G.~M.}\ \bibnamefont
  {Harry}},\ }\href {\doibase 10.1088/0264-9381/27/8/084006} {\bibfield
  {journal} {\bibinfo  {journal} {Classical and Quantum Gravity}\ }\textbf
  {\bibinfo {volume} {27}},\ \bibinfo {pages} {084006} (\bibinfo {year}
  {2010})}\BibitemShut {NoStop}%
\bibitem [{\citenamefont {Bondurant}\ and\ \citenamefont
  {Shapiro}(1984)}]{bondurant1984}%
  \BibitemOpen
  \bibfield  {author} {\bibinfo {author} {\bibfnamefont {R.}~\bibnamefont
  {Bondurant}}\ and\ \bibinfo {author} {\bibfnamefont {J.}~\bibnamefont
  {Shapiro}},\ }\href {\doibase 10.1103/PhysRevD.30.2548} {\bibfield  {journal}
  {\bibinfo  {journal} {Physical Review D}\ }\textbf {\bibinfo {volume} {30}},\
  \bibinfo {pages} {2548} (\bibinfo {year} {1984})}\BibitemShut {NoStop}%
\bibitem [{\citenamefont {Kimble}\ \emph {et~al.}(2001)\citenamefont {Kimble},
  \citenamefont {Levin}, \citenamefont {Matsko}, \citenamefont {Thorne},\ and\
  \citenamefont {Vyatchanin}}]{Kimble2001}%
  \BibitemOpen
  \bibfield  {author} {\bibinfo {author} {\bibfnamefont {H.}~\bibnamefont
  {Kimble}}, \bibinfo {author} {\bibfnamefont {Y.}~\bibnamefont {Levin}},
  \bibinfo {author} {\bibfnamefont {A.}~\bibnamefont {Matsko}}, \bibinfo
  {author} {\bibfnamefont {K.}~\bibnamefont {Thorne}}, \ and\ \bibinfo {author}
  {\bibfnamefont {S.}~\bibnamefont {Vyatchanin}},\ }\href {\doibase
  10.1103/PhysRevD.65.022002} {\bibfield  {journal} {\bibinfo  {journal}
  {Physical Review D}\ }\textbf {\bibinfo {volume} {65}},\ \bibinfo {pages}
  {022002} (\bibinfo {year} {2001})}\BibitemShut {NoStop}%
\bibitem [{\citenamefont {Bondurant}(1986)}]{Bondurant1986}%
  \BibitemOpen
  \bibfield  {author} {\bibinfo {author} {\bibfnamefont {R.}~\bibnamefont
  {Bondurant}},\ }\href {\doibase 10.1103/PhysRevA.34.3927} {\bibfield
  {journal} {\bibinfo  {journal} {Physical Review A}\ }\textbf {\bibinfo
  {volume} {34}},\ \bibinfo {pages} {3927} (\bibinfo {year}
  {1986})}\BibitemShut {NoStop}%
\bibitem [{\citenamefont {Briant}\ \emph {et~al.}(2003)\citenamefont {Briant},
  \citenamefont {Cerdonio}, \citenamefont {Conti}, \citenamefont {Heidmann},
  \citenamefont {Lobo},\ and\ \citenamefont {Pinard}}]{Briant2003}%
  \BibitemOpen
  \bibfield  {author} {\bibinfo {author} {\bibfnamefont {T.}~\bibnamefont
  {Briant}}, \bibinfo {author} {\bibfnamefont {M.}~\bibnamefont {Cerdonio}},
  \bibinfo {author} {\bibfnamefont {L.}~\bibnamefont {Conti}}, \bibinfo
  {author} {\bibfnamefont {A.}~\bibnamefont {Heidmann}}, \bibinfo {author}
  {\bibfnamefont {A.~E.}\ \bibnamefont {Lobo}}, \ and\ \bibinfo {author}
  {\bibfnamefont {M.}~\bibnamefont {Pinard}},\ }\href {\doibase
  10.1103/PhysRevD.68.102005} {\bibfield  {journal} {\bibinfo  {journal}
  {Physical Review D}\ }\textbf {\bibinfo {volume} {68}},\ \bibinfo {pages}
  {102005} (\bibinfo {year} {2003})}\BibitemShut {NoStop}%
\bibitem [{\citenamefont {Caniard}\ \emph {et~al.}(2007)\citenamefont
  {Caniard}, \citenamefont {Verlot}, \citenamefont {Briant}, \citenamefont
  {Cohadon},\ and\ \citenamefont {Heidmann}}]{Caniard2007}%
  \BibitemOpen
  \bibfield  {author} {\bibinfo {author} {\bibfnamefont {T.}~\bibnamefont
  {Caniard}}, \bibinfo {author} {\bibfnamefont {P.}~\bibnamefont {Verlot}},
  \bibinfo {author} {\bibfnamefont {T.}~\bibnamefont {Briant}}, \bibinfo
  {author} {\bibfnamefont {P.-F.}\ \bibnamefont {Cohadon}}, \ and\ \bibinfo
  {author} {\bibfnamefont {A.}~\bibnamefont {Heidmann}},\ }\href {\doibase
  10.1103/PhysRevLett.99.110801} {\bibfield  {journal} {\bibinfo  {journal}
  {Physical Review Letters}\ }\textbf {\bibinfo {volume} {99}},\ \bibinfo
  {pages} {110801} (\bibinfo {year} {2007})}\BibitemShut {NoStop}%
\bibitem [{\citenamefont {Chen}\ \emph {et~al.}(2010)\citenamefont {Chen},
  \citenamefont {Danilishin}, \citenamefont {Khalili},\ and\ \citenamefont
  {M\"{u}ller-Ebhardt}}]{Chen2010}%
  \BibitemOpen
  \bibfield  {author} {\bibinfo {author} {\bibfnamefont {Y.}~\bibnamefont
  {Chen}}, \bibinfo {author} {\bibfnamefont {S.~L.}\ \bibnamefont
  {Danilishin}}, \bibinfo {author} {\bibfnamefont {F.~Y.}\ \bibnamefont
  {Khalili}}, \ and\ \bibinfo {author} {\bibfnamefont {H.}~\bibnamefont
  {M\"{u}ller-Ebhardt}},\ }\href {\doibase 10.1007/s10714-010-1060-y}
  {\bibfield  {journal} {\bibinfo  {journal} {General Relativity and
  Gravitation}\ }\textbf {\bibinfo {volume} {43}},\ \bibinfo {pages} {671}
  (\bibinfo {year} {2010})}\BibitemShut {NoStop}%
\bibitem [{\citenamefont {Braginsky}\ and\ \citenamefont
  {Khalili}(1995)}]{Braginsky1995a}%
  \BibitemOpen
  \bibfield  {author} {\bibinfo {author} {\bibfnamefont {V.~B.}\ \bibnamefont
  {Braginsky}}\ and\ \bibinfo {author} {\bibfnamefont {F.~Y.}\ \bibnamefont
  {Khalili}},\ }\href
  {http://books.google.de/books/about/Quantum\_Measurement.html?id=HHKRzXIVdxcC\&pgis=1}
  {\emph {\bibinfo {title} {{Quantum Measurement}}}}\ (\bibinfo {year}
  {1995})\BibitemShut {NoStop}%
\bibitem [{\citenamefont {Braginsky}\ and\ \citenamefont
  {Vorontsov}(1975)}]{Braginsky1975}%
  \BibitemOpen
  \bibfield  {author} {\bibinfo {author} {\bibfnamefont {V.~B.}\ \bibnamefont
  {Braginsky}}\ and\ \bibinfo {author} {\bibfnamefont {Y.~I.}\ \bibnamefont
  {Vorontsov}},\ }\href@noop {} {\bibfield  {journal} {\bibinfo  {journal}
  {Sov. Phys. Usp.}\ }\textbf {\bibinfo {volume} {17}},\ \bibinfo {pages} {644}
  (\bibinfo {year} {1975})}\BibitemShut {NoStop}%
\bibitem [{\citenamefont {Thorne}\ \emph {et~al.}(1978)\citenamefont {Thorne},
  \citenamefont {Drever}, \citenamefont {Caves}, \citenamefont {Zimmermann},\
  and\ \citenamefont {Sandberg}}]{Thorne1978}%
  \BibitemOpen
  \bibfield  {author} {\bibinfo {author} {\bibfnamefont {K.}~\bibnamefont
  {Thorne}}, \bibinfo {author} {\bibfnamefont {R.}~\bibnamefont {Drever}},
  \bibinfo {author} {\bibfnamefont {C.}~\bibnamefont {Caves}}, \bibinfo
  {author} {\bibfnamefont {M.}~\bibnamefont {Zimmermann}}, \ and\ \bibinfo
  {author} {\bibfnamefont {V.}~\bibnamefont {Sandberg}},\ }\href
  {http://link.aps.org/doi/10.1103/PhysRevLett.40.667} {\bibfield  {journal}
  {\bibinfo  {journal} {Physical Review Letters}\ }\textbf {\bibinfo {volume}
  {40}},\ \bibinfo {pages} {667} (\bibinfo {year} {1978})}\BibitemShut
  {NoStop}%
\bibitem [{\citenamefont {Braginsky}\ \emph {et~al.}(1980)\citenamefont
  {Braginsky}, \citenamefont {Vorontsov},\ and\ \citenamefont
  {Thorne}}]{Braginsky1980a}%
  \BibitemOpen
  \bibfield  {author} {\bibinfo {author} {\bibfnamefont {V.~B.}\ \bibnamefont
  {Braginsky}}, \bibinfo {author} {\bibfnamefont {Y.~I.}\ \bibnamefont
  {Vorontsov}}, \ and\ \bibinfo {author} {\bibfnamefont {K.~S.}\ \bibnamefont
  {Thorne}},\ }\href {http://www.sciencemag.org/content/209/4456/547}
  {\bibfield  {journal} {\bibinfo  {journal} {Science (New York, N.Y.)}\
  }\textbf {\bibinfo {volume} {209}},\ \bibinfo {pages} {547} (\bibinfo {year}
  {1980})}\BibitemShut {NoStop}%
\bibitem [{\citenamefont {Clerk}\ \emph {et~al.}(2008)\citenamefont {Clerk},
  \citenamefont {Marquardt},\ and\ \citenamefont {Jacobs}}]{Clerk2008}%
  \BibitemOpen
  \bibfield  {author} {\bibinfo {author} {\bibfnamefont {A.~A.}\ \bibnamefont
  {Clerk}}, \bibinfo {author} {\bibfnamefont {F.}~\bibnamefont {Marquardt}}, \
  and\ \bibinfo {author} {\bibfnamefont {K.}~\bibnamefont {Jacobs}},\ }\href
  {http://iopscience.iop.org/1367-2630/10/9/095010/fulltext/} {\bibfield
  {journal} {\bibinfo  {journal} {New Journal of Physics}\ }\textbf {\bibinfo
  {volume} {10}},\ \bibinfo {pages} {095010} (\bibinfo {year}
  {2008})}\BibitemShut {NoStop}%
\bibitem [{\citenamefont {Suh}\ \emph {et~al.}(2013)\citenamefont {Suh},
  \citenamefont {Weinstein}, \citenamefont {Lei}, \citenamefont {Wollman},
  \citenamefont {Steinke}, \citenamefont {Meystre}, \citenamefont {Clerk},\
  and\ \citenamefont {Schwab}}]{Suh2013a}%
  \BibitemOpen
  \bibfield  {author} {\bibinfo {author} {\bibfnamefont {J.}~\bibnamefont
  {Suh}}, \bibinfo {author} {\bibfnamefont {A.~J.}\ \bibnamefont {Weinstein}},
  \bibinfo {author} {\bibfnamefont {C.~U.}\ \bibnamefont {Lei}}, \bibinfo
  {author} {\bibfnamefont {E.~E.}\ \bibnamefont {Wollman}}, \bibinfo {author}
  {\bibfnamefont {S.~K.}\ \bibnamefont {Steinke}}, \bibinfo {author}
  {\bibfnamefont {P.}~\bibnamefont {Meystre}}, \bibinfo {author} {\bibfnamefont
  {A.~A.}\ \bibnamefont {Clerk}}, \ and\ \bibinfo {author} {\bibfnamefont
  {K.~C.}\ \bibnamefont {Schwab}},\ }\href {http://arxiv.org/abs/1312.4084} {\
  (\bibinfo {year} {2013})},\ \Eprint {http://arxiv.org/abs/1312.4084}
  {arXiv:1312.4084} \BibitemShut {NoStop}%
\bibitem [{\citenamefont {Chelkowski}\ \emph {et~al.}(2005)\citenamefont
  {Chelkowski}, \citenamefont {Vahlbruch}, \citenamefont {Hage}, \citenamefont
  {Franzen}, \citenamefont {Lastzka}, \citenamefont {Danzmann},\ and\
  \citenamefont {Schnabel}}]{Chelkowski2005}%
  \BibitemOpen
  \bibfield  {author} {\bibinfo {author} {\bibfnamefont {S.}~\bibnamefont
  {Chelkowski}}, \bibinfo {author} {\bibfnamefont {H.}~\bibnamefont
  {Vahlbruch}}, \bibinfo {author} {\bibfnamefont {B.}~\bibnamefont {Hage}},
  \bibinfo {author} {\bibfnamefont {A.}~\bibnamefont {Franzen}}, \bibinfo
  {author} {\bibfnamefont {N.}~\bibnamefont {Lastzka}}, \bibinfo {author}
  {\bibfnamefont {K.}~\bibnamefont {Danzmann}}, \ and\ \bibinfo {author}
  {\bibfnamefont {R.}~\bibnamefont {Schnabel}},\ }\href {\doibase
  10.1103/PhysRevA.71.013806} {\bibfield  {journal} {\bibinfo  {journal}
  {Physical Review A}\ }\textbf {\bibinfo {volume} {71}},\ \bibinfo {pages}
  {013806} (\bibinfo {year} {2005})}\BibitemShut {NoStop}%
\bibitem [{\citenamefont {Mow-Lowry}\ \emph {et~al.}(2004)\citenamefont
  {Mow-Lowry}, \citenamefont {Sheard}, \citenamefont {Gray}, \citenamefont
  {McClelland},\ and\ \citenamefont {Whitcomb}}]{Mow-Lowry2004}%
  \BibitemOpen
  \bibfield  {author} {\bibinfo {author} {\bibfnamefont {C.}~\bibnamefont
  {Mow-Lowry}}, \bibinfo {author} {\bibfnamefont {B.}~\bibnamefont {Sheard}},
  \bibinfo {author} {\bibfnamefont {M.}~\bibnamefont {Gray}}, \bibinfo {author}
  {\bibfnamefont {D.}~\bibnamefont {McClelland}}, \ and\ \bibinfo {author}
  {\bibfnamefont {S.}~\bibnamefont {Whitcomb}},\ }\href {\doibase
  10.1103/PhysRevLett.92.161102} {\bibfield  {journal} {\bibinfo  {journal}
  {Physical Review Letters}\ }\textbf {\bibinfo {volume} {92}},\ \bibinfo
  {pages} {161102} (\bibinfo {year} {2004})}\BibitemShut {NoStop}%
\bibitem [{\citenamefont {Sheard}\ \emph {et~al.}(2004)\citenamefont {Sheard},
  \citenamefont {Gray}, \citenamefont {Mow-Lowry}, \citenamefont {McClelland},\
  and\ \citenamefont {Whitcomb}}]{Sheard2004}%
  \BibitemOpen
  \bibfield  {author} {\bibinfo {author} {\bibfnamefont {B.}~\bibnamefont
  {Sheard}}, \bibinfo {author} {\bibfnamefont {M.}~\bibnamefont {Gray}},
  \bibinfo {author} {\bibfnamefont {C.}~\bibnamefont {Mow-Lowry}}, \bibinfo
  {author} {\bibfnamefont {D.}~\bibnamefont {McClelland}}, \ and\ \bibinfo
  {author} {\bibfnamefont {S.}~\bibnamefont {Whitcomb}},\ }\href {\doibase
  10.1103/PhysRevA.69.051801} {\bibfield  {journal} {\bibinfo  {journal}
  {Physical Review A}\ }\textbf {\bibinfo {volume} {69}},\ \bibinfo {pages}
  {051801} (\bibinfo {year} {2004})}\BibitemShut {NoStop}%
\bibitem [{\citenamefont {Tsang}\ and\ \citenamefont
  {Caves}(2010)}]{Tsang2010}%
  \BibitemOpen
  \bibfield  {author} {\bibinfo {author} {\bibfnamefont {M.}~\bibnamefont
  {Tsang}}\ and\ \bibinfo {author} {\bibfnamefont {C.}~\bibnamefont {Caves}},\
  }\href {\doibase 10.1103/PhysRevLett.105.123601} {\bibfield  {journal}
  {\bibinfo  {journal} {Physical Review Letters}\ }\textbf {\bibinfo {volume}
  {105}} (\bibinfo {year} {2010}),\ 10.1103/PhysRevLett.105.123601}\BibitemShut
  {NoStop}%
\bibitem [{\citenamefont {Julsgaard}\ \emph {et~al.}(2001)\citenamefont
  {Julsgaard}, \citenamefont {Kozhekin},\ and\ \citenamefont
  {Polzik}}]{Julsgaard2001}%
  \BibitemOpen
  \bibfield  {author} {\bibinfo {author} {\bibfnamefont {B.}~\bibnamefont
  {Julsgaard}}, \bibinfo {author} {\bibfnamefont {A.}~\bibnamefont {Kozhekin}},
  \ and\ \bibinfo {author} {\bibfnamefont {E.~S.}\ \bibnamefont {Polzik}},\
  }\href {\doibase 10.1038/35096524} {\bibfield  {journal} {\bibinfo  {journal}
  {Nature}\ }\textbf {\bibinfo {volume} {413}},\ \bibinfo {pages} {400}
  (\bibinfo {year} {2001})}\BibitemShut {NoStop}%
\bibitem [{\citenamefont {Hammerer}\ \emph {et~al.}(2009)\citenamefont
  {Hammerer}, \citenamefont {Aspelmeyer}, \citenamefont {Polzik},\ and\
  \citenamefont {Zoller}}]{Hammerer2009}%
  \BibitemOpen
  \bibfield  {author} {\bibinfo {author} {\bibfnamefont {K.}~\bibnamefont
  {Hammerer}}, \bibinfo {author} {\bibfnamefont {M.}~\bibnamefont
  {Aspelmeyer}}, \bibinfo {author} {\bibfnamefont {E.}~\bibnamefont {Polzik}},
  \ and\ \bibinfo {author} {\bibfnamefont {P.}~\bibnamefont {Zoller}},\ }\href
  {\doibase 10.1103/PhysRevLett.102.020501} {\bibfield  {journal} {\bibinfo
  {journal} {Physical Review Letters}\ }\textbf {\bibinfo {volume} {102}}
  (\bibinfo {year} {2009}),\ 10.1103/PhysRevLett.102.020501}\BibitemShut
  {NoStop}%
\bibitem [{\citenamefont {Wasilewski}\ \emph {et~al.}(2010)\citenamefont
  {Wasilewski}, \citenamefont {Jensen}, \citenamefont {Krauter}, \citenamefont
  {Renema}, \citenamefont {Balabas},\ and\ \citenamefont
  {Polzik}}]{Wasilewski2010}%
  \BibitemOpen
  \bibfield  {author} {\bibinfo {author} {\bibfnamefont {W.}~\bibnamefont
  {Wasilewski}}, \bibinfo {author} {\bibfnamefont {K.}~\bibnamefont {Jensen}},
  \bibinfo {author} {\bibfnamefont {H.}~\bibnamefont {Krauter}}, \bibinfo
  {author} {\bibfnamefont {J.~J.}\ \bibnamefont {Renema}}, \bibinfo {author}
  {\bibfnamefont {M.~V.}\ \bibnamefont {Balabas}}, \ and\ \bibinfo {author}
  {\bibfnamefont {E.~S.}\ \bibnamefont {Polzik}},\ }\href {\doibase
  10.1103/PhysRevLett.104.133601} {\bibfield  {journal} {\bibinfo  {journal}
  {Physical Review Letters}\ }\textbf {\bibinfo {volume} {104}} (\bibinfo
  {year} {2010}),\ 10.1103/PhysRevLett.104.133601}\BibitemShut {NoStop}%
\bibitem [{\citenamefont {Zhang}\ \emph {et~al.}(2013)\citenamefont {Zhang},
  \citenamefont {Meystre},\ and\ \citenamefont {Zhang}}]{Zhang2013a}%
  \BibitemOpen
  \bibfield  {author} {\bibinfo {author} {\bibfnamefont {K.}~\bibnamefont
  {Zhang}}, \bibinfo {author} {\bibfnamefont {P.}~\bibnamefont {Meystre}}, \
  and\ \bibinfo {author} {\bibfnamefont {W.}~\bibnamefont {Zhang}},\ }\href
  {http://link.aps.org/doi/10.1103/PhysRevA.88.043632} {\bibfield  {journal}
  {\bibinfo  {journal} {Physical Review A}\ }\textbf {\bibinfo {volume} {88}},\
  \bibinfo {pages} {043632} (\bibinfo {year} {2013})}\BibitemShut {NoStop}%
\bibitem [{\citenamefont {Tsang}\ and\ \citenamefont
  {Caves}(2012)}]{Tsang2012}%
  \BibitemOpen
  \bibfield  {author} {\bibinfo {author} {\bibfnamefont {M.}~\bibnamefont
  {Tsang}}\ and\ \bibinfo {author} {\bibfnamefont {C.~M.}\ \bibnamefont
  {Caves}},\ }\href {\doibase 10.1103/PhysRevX.2.031016} {\bibfield  {journal}
  {\bibinfo  {journal} {Physical Review X}\ }\textbf {\bibinfo {volume} {2}},\
  \bibinfo {pages} {031016} (\bibinfo {year} {2012})}\BibitemShut {NoStop}%
\bibitem [{\citenamefont {Gough}\ and\ \citenamefont
  {James}(2009)}]{Gough2009}%
  \BibitemOpen
  \bibfield  {author} {\bibinfo {author} {\bibfnamefont {J.}~\bibnamefont
  {Gough}}\ and\ \bibinfo {author} {\bibfnamefont {M.}~\bibnamefont {James}},\
  }\href
  {http://ieeexplore.ieee.org/lpdocs/epic03/wrapper.htm?arnumber=5286277}
  {\bibfield  {journal} {\bibinfo  {journal} {IEEE Transactions on Automatic
  Control}\ }\textbf {\bibinfo {volume} {54}},\ \bibinfo {pages} {2530}
  (\bibinfo {year} {2009})}\BibitemShut {NoStop}%
\bibitem [{Note1()}]{Note1}%
  \BibitemOpen
  \bibinfo {note} {In the high temperature limit considered here $\protect
  \mathaccentV {bar}016{n}\gg 1$.}\BibitemShut {Stop}%
\bibitem [{\citenamefont {Gardiner}\ and\ \citenamefont
  {Zoller}(2004)}]{Gardiner2004}%
  \BibitemOpen
  \bibfield  {author} {\bibinfo {author} {\bibfnamefont {C.}~\bibnamefont
  {Gardiner}}\ and\ \bibinfo {author} {\bibfnamefont {P.}~\bibnamefont
  {Zoller}},\ }\href
  {http://www.amazon.com/Quantum-Noise-Non-Markovian-Applications-Synergetics/dp/3540223010}
  {\emph {\bibinfo {title} {{Quantum Noise: A Handbook of Markovian and
  Non-Markovian Quantum Stochastic Methods with Applications to Quantum Optics
  (Springer Series in Synergetics)}}}}\ (\bibinfo  {publisher} {Springer},\
  \bibinfo {year} {2004})\BibitemShut {NoStop}%
\bibitem [{\citenamefont {Byer}\ and\ \citenamefont {Herbst}(1977)}]{Byer1977}%
  \BibitemOpen
  \bibfield  {author} {\bibinfo {author} {\bibfnamefont {R.~L.}\ \bibnamefont
  {Byer}}\ and\ \bibinfo {author} {\bibfnamefont {R.~L.}\ \bibnamefont
  {Herbst}},\ }in\ \href {\doibase 10.1007/3540079459\_9} {\emph {\bibinfo
  {booktitle} {Nonlinear Infrared Generation}}}\ (\bibinfo {year} {1977})\ pp.\
  \bibinfo {pages} {81--137}\BibitemShut {NoStop}%
\end{thebibliography}%

\end{document}